\documentclass[longauth]{aa}  

\usepackage{graphicx}
\usepackage{txfonts}
\usepackage{hyperref}
\hypersetup{
    colorlinks=true,
    linkcolor=blue,
    citecolor=blue,
    filecolor=magenta,      
    urlcolor=cyan,
    }
\usepackage{textgreek}
\usepackage{booktabs}

\newcommand{\lya}{Ly\textalpha}
\newcommand{\ha}{H\textalpha}
\newcommand{\hb}{H\textbeta}
\newcommand{\hg}{H\textgamma}
\newcommand{\oiii}{[O\,\textsc{iii}]}
\newcommand{\oii}{[O\,\textsc{ii}]}
\newcommand{\hii}{H\,\textsc{ii}}

\newcommand{\xiion}{$\xi_\mathrm{ion}$}
\newcommand{\fesc}{$f_\mathrm{esc}$}

\newcommand{\flux}{erg\,s$^{-1}$\,cm$^{-2}$}

\newcommand{\kms}{km\,s$^{-1}$}

\begin{document}

\title{JADES: Discovery of extremely high equivalent width Lyman-alpha emission from a faint galaxy within an ionized bubble at $z=7.3$}
\titlerunning{Discovery of an extreme LAE at $z=7.3$}

\author{
Aayush Saxena\inst{1,2}\thanks{E-mail: aayush.saxena@physics.ox.ac.uk}
\and Brant E. Robertson\inst{3}
\and Andrew J. Bunker\inst{1}
\and Ryan Endsley\inst{4}
\and Alex J. Cameron\inst{1}
\and Stephane Charlot\inst{5}
\and Charlotte Simmonds\inst{6,7}
\and Sandro Tacchella\inst{6,7}
\and Joris Witstok\inst{6,7}
\and Chris Willott\inst{8}
\and Stefano Carniani\inst{9}
\and Emma Curtis-Lake\inst{10}
\and Pierre Ferruit\inst{11}
\and Peter Jakobsen\inst{12,13}
\and Santiago Arribas\inst{14}
\and Jacopo Chevallard\inst{1}
\and Mirko Curti\inst{6,7,15}
\and Francesco D'Eugenio\inst{6,7}
\and Anna De Graaff\inst{16}
\and Gareth C. Jones\inst{1}
\and Tobias J. Looser\inst{6,7}
\and Michael V. Maseda\inst{17}
\and Tim Rawle\inst{18}
\and Hans-Walter Rix\inst{16}
\and Bruno Rodr\'{i}guez Del Pino\inst{14}
\and Renske Smit\inst{19}
\and Hannah \"{U}bler\inst{6,7}
\and Daniel J. Eisenstein\inst{20}
\and Kevin Hainline\inst{21}
\and Ryan Hausen\inst{22}
\and Benjamin D. Johnson\inst{20}
\and Marcia Rieke\inst{21}
\and Christina C. Williams\inst{23}
\and Christopher N. A. Willmer\inst{21}
\and William M. Baker\inst{6,7}
\and Rachana Bhatawdekar\inst{11,24}
\and Rebecca Bowler\inst{25}
\and Kristan Boyett\inst{26,27}
\and Zuyi Chen\inst{21}
\and Eiichi Egami\inst{21}
\and Zhiyuan Ji\inst{21}
\and Nimisha Kumari\inst{28}
\and Erica Nelson\inst{29}
\and Michele Perna\inst{14}
\and Lester Sandles\inst{6,7}
\and Jan Scholtz\inst{6,7,2}
\and Irene Shivaei\inst{21}
}

\institute{
Department of Physics, University of Oxford, Denys Wilkinson Building, Keble Road, Oxford OX1 3RH, UK
\and Department of Physics and Astronomy, University College London, Gower Street, London WC1E 6BT, UK
\and Department of Astronomy and Astrophysics, University of California, Santa Cruz, 1156 High Street, Santa Cruz, CA 95064, USA
\and Department of Astronomy, University of Texas, Austin, TX 78712, USA
\and Sorbonne Universit\'e, CNRS, UMR 7095, Institut d'Astrophysique de Paris, 98 bis bd Arago, 75014 Paris, France
\and Kavli Institute for Cosmology, University of Cambridge, Madingley Road, Cambridge CB3 0HA, UK
\and Cavendish Laboratory, University of Cambridge, 19 JJ Thomson Avenue, Cambridge CB3 0HE, UK
\and NRC Herzberg, 5071 West Saanich Rd, Victoria, BC V9E 2E7, Canada
\and Scuola Normale Superiore, Piazza dei Cavalieri 7, I-56126 Pisa, Italy
\and center for Astrophysics Research, Department of Physics, Astronomy and Mathematics, University of Hertfordshire, Hatfield AL10 9AB, UK
\and European Space Agency, European Space Astronomy center, Camino Bajo del Castillo s/n, 28692 Villafranca del Castillo, Madrid, Spain
\and Cosmic Dawn Center (DAWN), Copenhagen, Denmark
\and Niels Bohr Institute, University of Copenhagen, Jagtvej 128, DK-2200, Copenhagen, Denmark
\and Centro de Astrobiolog\'{i}a (CAB), CSIC-INTA, Cra. de Ajalvir Km.~4, 28850 Torrej\'{o}n de Ardoz, Madrid, Spain
\and European Southern Observatory, Karl-Schwarzschild-Strasse 2, D-85748 Garching bei Muenchen, Germany
\and Max-Planck-Institut f\"ur Astronomie, K\"onigstuhl 17, D-69117, Heidelberg, Germany
\and Department of Astronomy, University of Wisconsin-Madison, 475 N. Charter St., Madison, WI 53706, USA
\and European Space Agency, Space Telescope Science Institute, Baltimore, Maryland, US
\and Astrophysics Research Institute, Liverpool John Moores University, 146 Brownlow Hill, Liverpool L3 5RF, UK
\and Centre for Astrophysics $|$ Harvard \& Smithsonian, 60 Garden St., Cambridge MA 02138 USA
\and Steward Observatory, University of Arizona, 933 N. Cherry Ave., Tucson, AZ 85721, USA
\and Department of Physics and Astronomy, The Johns Hopkins University, 3400 N. Charles St., Baltimore, MD 21218
\and NSF's National Optical-Infrared Astronomy Research Laboratory, 950 North Cherry Avenue, Tucson, AZ 85719, USA
\and European Space Agency, ESA/ESTEC, Keplerlaan 1, 2201 AZ Noordwijk, NL
\and Jodrell Bank Centre for Astrophysics, Department of Physics and Astronomy, School of Natural Sciences, The University of Manchester, Manchester, M13 9PL, UK
\and School of Physics, University of Melbourne, Parkville 3010, VIC, Australia
\and ARC Centre of Excellence for All Sky Astrophysics in 3 Dimensions (ASTRO 3D), Australia
\and AURA for European Space Agency, Space Telescope Science Institute, 3700 San Martin Drive. Baltimore, MD, 21210
\and Department for Astrophysical and Planetary Science, University of Colorado, Boulder, CO 80309, USA
}

\authorrunning{A. Saxena et al.}

\date{Accepted: 29 August 2023}

%\abstract{}{}{}{}{} 
% 5 {} token are mandatory

\abstract{We report the discovery of a remarkable Ly$\alpha$ emitting galaxy at $z=7.2782$, JADES-GS+53.16746-27.7720 (shortened to JADES-GS-z7-LA), with rest-frame equivalent width, EW$_0$(Ly$\alpha$) $=388.0 \pm 88.8$ \AA\ and UV magnitude $-17.0$. The spectroscopic redshift is confirmed via rest-frame optical lines [O\,\textsc{ii}], H$\beta$ and [O\,\textsc{iii}] in its JWST/NIRSpec Micro-Shutter Assembly (MSA) spectrum. The Ly$\alpha$ line is detected in both lower resolution ($R\sim100$) PRISM as well as medium resolution ($R\sim1000$) G140M grating spectra. The line spread function-deconvolved Ly$\alpha$ full width at half maximum in the grating is $383.9 \pm 56.2$\,km s$^{-1}$ and the Ly$\alpha$ velocity offset compared to the systemic redshift is $113.3 \pm 80.0$\,km s$^{-1}$, indicative of very little neutral gas or dust within the galaxy. We estimate the Ly$\alpha$ escape fraction to be $>70\%$. JADES-GS-z7-LA has a [O\,\textsc{iii}]/[O\,\textsc{ii}] ratio (O32) of $11.1 \pm 2.2$ and a ([O\,\textsc{iii}]+[O\,\textsc{ii}])/H$\beta$ ratio (R23) of $11.2\pm2.6$, consistent with low metallicity and high ionization parameters. Deep NIRCam imaging also revealed a close companion source (separated by $0.23''$), which exhibits similar photometry to that of JADES-GS-z7-LA, with a photometric excess in the F410M NIRCam image consistent with [O\,\textsc{iii}]+H$\beta$ emission at the same redshift. The spectral energy distribution of JADES-GS-z7-LA indicates a ``bursty'' star formation history, with a low stellar mass of $\approx 10^{7}$ M$_\odot$. Assuming that the Ly$\alpha$ transmission through the intergalactic medium is the same as its measured escape fraction, an ionized region of size $>1.5$ pMpc is needed to explain the high Ly$\alpha$ EW and low velocity offset compared to systemic seen in JADES-GS-z7-LA. Owing to its UV-faintness, we show that it is incapable of single-handedly ionizing a region large enough to explain its Ly$\alpha$ emission. Therefore, we suggest that JADES-GS-z7-LA (and possibly the companion source) may be a part of a larger overdensity, presenting direct evidence of overlapping ionized bubbles at $z>7$.}


\maketitle

\section{Introduction} 
\label{sec:intro}

Understanding the key drivers of cosmic reionization - an important phase transition in the Universe's lifetime whereby the intergalactic medium (IGM) transformed from neutral to completely ionized by $z\sim6$ - is an important challenge in observational astronomy \citep[see][for a recent review]{rob22a}. This requires good estimates on the rate of the hydrogen ionizing Lyman continuum (LyC; $\lambda_0 < 912$\,\AA) photons that are produced within galaxies, as well as the fraction of these that are able to escape from galaxies, contributing toward reionization of the IGM \citep{day18}.

Although early data from the \emph{James Webb Space Telescope (JWST)} have already provided unprecedented insights into the interstellar medium (ISM) conditions and ionizing photon production in reionization era galaxies at $z>6$ \citep{are22, curtis23, rob23, end23, cur23, kat23b, tru23, tac23a, sun23, san23, fuj23, cam23}, placing direct constraints on the escape fraction of LyC photons (\fesc) is impossible at $z>6$ because of the increasing neutrality of the IGM that absorbs virtually all LyC photons along the line of sight \citep[e.g.,][]{ino14}. Even so, there are indirect indicators of LyC \fesc\ that could potentially be used to estimate ionizing photon escape using \emph{JWST} data \citep[e.g.,][]{zac13, top22, mas23}.

However, as individual galaxies or groups of galaxies begin to reionize their immediate surroundings, the ionized ``bubbles'' will eventually grow large enough to allow the \lya\ photons to escape along the line of sight without considerable absorption, so long as they have been sufficiently redshifted out of \lya\ resonance \citep{mir98, fur06, mason20, tra23}. By then studying the evolution of the physical properties of \lya-emitting galaxies (LAEs) as a function of redshift \citep[e.g.,][]{mal06, pen11, pen14, zit15, fur16, sta17, mas19a, til20, jun20, whi20, hu21, cas18, cas22, end22a, nin22}, accurate constraints can be placed on the evolution of the neutral fraction of the IGM across redshifts by carefully modeling \lya\ emission and comparing it with models of inhomogeneous reionization \citep[e.g.,][]{mes15, mas18, mason20, mor21, bol22, qin22, mat22, tra23}. The spatial distribution and associations of these \lya\ emitting galaxies can give further insights into whether reionization proceeds uniformly across the Universe or in a patchy manner \citep[e.g.,][]{lar22, leo22, tan23}.

Before \emph{JWST}, \lya\ observations were mostly attempted from the ground \citep[e.g.,][]{van11, pen14, oes15, fur16, sta17, lap17, roberts23, nin22}, and as a result, most were restricted to some of the brightest known galaxies at $z>6$.  The bright LAEs at $z>6$ are good beacons of ionized regions in the Universe, as they would have undergone intense star formation activity that would eventually have led to the inflation of a bubble large enough for \lya\ emission to pass through \citep[e.g.,][]{mad99}. However, such galaxies are rare, and according to certain models of reionization, the bulk of the photons ionizing the IGM are expected to come from the more numerous fainter, lower mass galaxies at $z>6$ \citep[e.g.,][]{rob15, fin19}. Therefore, it is vital to begin assembling a picture of \lya\ emission and transmission from fainter UV galaxies to better understand their role in driving reionization, and to uncover additional independent sightlines leading to a more complete picture of the spatial and temporal evolution of reionization.

Observations around lensing clusters have already helped push \lya\ detections from the ground to fainter UV magnitudes \citep[e.g.,][]{hoa19, ful20, bol22}. Spectroscopic follow-up of faint high redshift galaxy candidates in well-studied extragalactic fields with \emph{JWST} now offers an excellent opportunity to investigate the presence of \lya\ emission in such sources at $z>6$. In this paper, we leverage the power of the \emph{JWST} Advanced Deep Extragalactic Survey (JADES; \citealt{eis23}) and present the discovery of a very faint galaxy spectroscopically confirmed at $z=7.278$. The aim of this paper is to constrain both the ISM conditions and ionization properties of the \lya\ emitting galaxy as well as characterize the state of the IGM around it, making inferences on its implications for the reionization of the Universe.

The layout of this paper is as follows: Section \ref{sec:data} describes the \emph{JWST} data used in this study. Section \ref{sec:lae} presents details about the observed properties of the remarkable LAE at $z=7.278$, JADES-GS+53.16746-27.77201, shortened to JADES-GS-z7-LA, which is the main focus of this paper. Section \ref{sec:SED} contains details of modeling the spectral energy distribution (SED), leading to stronger constraints on the physical parameters of the galaxy. Section \ref{sec:reionization} discusses the implications for the reionization of the Universe from these observations. The conclusions are presented in Section \ref{sec:conclusions}.

Throughout this paper, we use the \citet{planck} cosmology. Magnitudes are in the AB system \citep{oke83} and all distances used are proper distances, unless otherwise stated.

\section{Data} 
\label{sec:data}

JADES-GS+53.16746-27.77201 was first reported in \citet{mcl13} and \citet{sch13} as a high redshift galaxy candidate using \emph{HST} imaging in the HUDF with ID UDF12-4019-6190 and a photometric redshift of $z_{\rm{phot}}=6.9$. Being a high redshift candidate, this object was included in JADES NIRSpec and NIRCam observations that are described below.

The JWST observations used in this study are part of JADES, which is a collaboration between the Near-Infrared Camera (NIRCam; \citealt{rie23a}) and Near-Infrared Spectrograph (NIRSpec; \citealt{fer22}) Instrument Science teams with an aim of using over 750 hours of guaranteed time observations (GTO) to study the evolution of galaxies in the Great Observatories Origins Deep Survey (GOODS)-South and GOODS-North fields. We describe the NIRSpec and NIRCam observations and data reduction steps that led to the discovery of JADES-GS-z7-LA below.

\subsection{NIRSpec spectroscopy}
The NIRSpec observations used here are part of the GTO program ID: 1210 (PI: L\"utzgendorf) in GOODS-S, using the Micro-Shutter Assembly (MSA; \citealt{jak22, fer22}) centered near the Hubble Ultra Deep Field (HUDF), obtained between 22 October and 25 October 2022 over 3 visits. Each visit had a total of 33,613 seconds of exposure in the PRISM/CLEAR setup, which gives wavelength coverage in the range $0.6-5.3$ $\mu$m with a spectral resolution of $R\sim100$ \citep{bok23}, and 8,403 seconds integration in each of G140M/F070LP, G235M/F170LP, G395M/F290LP, and G395H/F290LP filter and grating setups. We refer the readers to \citet{bun23b} for further details about the observational setup. 

The targets for spectroscopy were selected from existing deep \textit{HST} imaging and catalogues in the field. Candidate high redshift galaxies with photometric redshifts $z>5.7$, identified via the classic photometric \lq drop-out' technique \citep[e.g.,][]{ste96}, whereby the Lyman break in the spectrum of a galaxy is captured in adjacent broad-band filters, were assigned higher priorities. Full details of the target selection and priority classes can be found in \citet{bun23b}.

As also mentioned in \citet{cam23} and \citet{bun23b}, each of the three visits were slightly offset ($<1$'') and had a unique MSA configuration, with target allocation optimized to achieve coverage of the highest priority targets across multiple visits. JADES-GS-z7-LA was covered in 2 out of the 3 visits, which meant that it had a total exposure time of 18.7 hours in PRISM/CLEAR and 4.7 hours in each of the mid/high resolution filter/grating setups. For each visit, a three-point nodding pattern was used for background subtraction.

The data reduction was carried out using pipelines developed by the ESA NIRSpec Science Operations Team (SOT) and the NIRSpec GTO Team \citep{fer22}, which are customized versions of the official data reduction pipeline supplied by STScI. Some of the  main data reduction steps implemented by the pipeline are pixel-level background subtraction, pixel-to-pixel flat-field correction, absolute flux calibration, slit-loss correction, and eventually 2D and 1D spectra extraction and co-addition. In this version of the reduction, the final 1D spectra are not extracted from the 2D spectra, but result from the weighted averaging of 1D spectra from all integrations \citep[see][]{bun23b}. Due to the compact size on the sky ($\approx0.1''$) of JADES-GS-z7-LA, slit-loss corrections were applied by modeling it as point-like source. A nominal 3-pixel extraction aperture was used to produce the co-added 1D spectra. A detailed description of the data reduction and spectral extraction methods is given in \citet{curtis23} and \citet{cam23}.

\subsection{NIRCam imaging}
The  NIRCam data used in this study are part of the GTO program ID: 1180 (PI: Eisenstein) in GOODS-S, obtained between 29 September and 5 October 2022. A detailed description of the observations and data reduction can be found in \citet{rob23} and \citet{tac23b}, but here we briefly describe the methodology to obtain NIRCam images.

Four pointings centered at or around the HUDF were used to produce a $4.4 \times 6$ arcminute mosaic. The filters used for these observations were F090W, F115W, F150W and F200W in the short-wavelength (SW) channel, and F277W, F335M, F356W, F410M and F444W in the long-wavelength (LW) channel. Individual exposures of 1375 seconds were used in a nine-point dither pattern in all four pointings.

The JWST Calibration Pipeline v1.8.1 was used for data reduction, where the main reduction steps involved in the production of Stage-1 products were dark subtraction, distortion correction, bad-pixel masking, bias subtraction, and `snowball' removal \citep{bir22}. The Stage-2 processing included flat-fielding and flux calibration. Astrometric alignment was then performed using a custom version of JWST TweakReg. Full details of the NIRCam data reduction and mosaicing will be presented in a forthcoming paper (Tachella et al. in prep) and more details of the basic reduction steps can already be found in \citet{rob23} and \citet{tac23b}.

\begin{figure*}
    \centering
    \includegraphics[width=\textwidth]{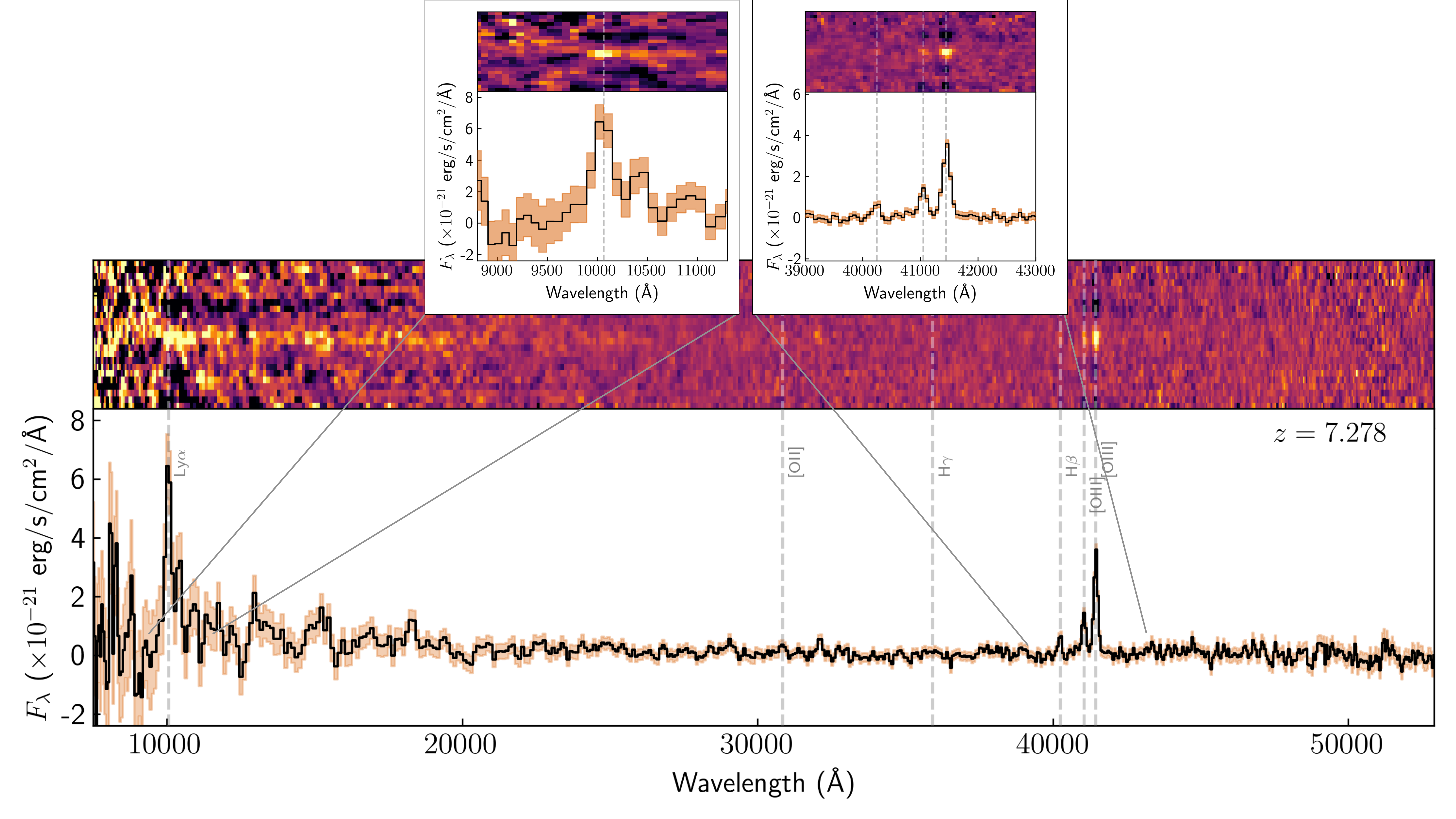}
    \caption{NIRSpec PRISM (R100) spectrum of JADES-GS-z7-LA at $z=7.278$, showing both 1D and 2D spectra. The error spectrum has been shown as the orange-shaded region. We additionally show zoom-ins around the \lya\ line (top left) and the \hb, \oiii$\lambda4959$ and \oiii$\lambda5007$ lines (top right). At least in the PRISM spectrum, the \lya\ emission appears at the systemic redshift, indicative of the presence of \lya\ (and potentially LyC) escape channels. We caution, however, that at the bluest end of the spectrograph the spectral resolution is much lower than the nominal $R\sim100$, degrading all the way down to $R\sim30$ \citep{jak22}.}
    \label{fig:R100_spec}
\end{figure*}

\section{A strong Lyman-alpha emitter at $\mathrm{z}\approx 7.3$} 
\label{sec:lae}
As mentioned earlier, in this paper we focus on JADES-GS+53.16746-27.77201, shortened to JADES-GS-z7-LA. We show the combined 1D and 2D spectra in Figure \ref{fig:R100_spec}, with zoom-ins around the wavelengths covering \lya\ and \oiii+\hb\ lines. Below we present both spectroscopic as well as photometric measurements for this remarkable LAE, placing constraints on dust and the star formation rate based on spectroscopic observations. We also present the discovery of a previously unknown Lyman break galaxy very close to JADES-GS-z7-LA.

\subsection{Spectroscopic redshift}
\begin{figure}
    \centering
    \includegraphics[width=\linewidth]{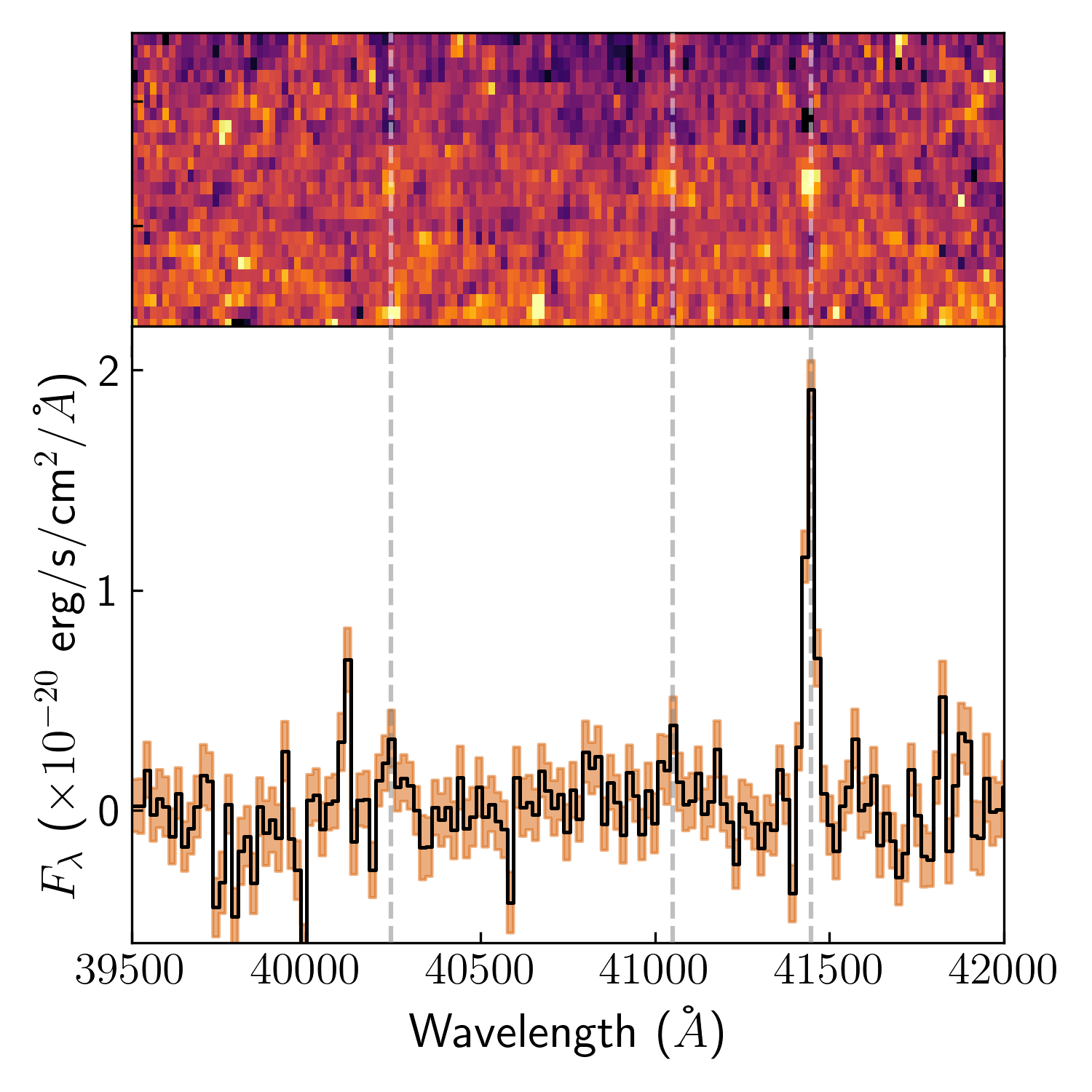}
    \caption{Robust detection of the \oiii\,$\lambda5007$ emission line in the $R\sim1000$ G395M grating spectrum, which was used to derive an accurate systemic redshift of $z=7.2782 \pm 0.0003$. This redshift is consistent with the weaker detections of the \oiii\,$\lambda4959$ and \hb\ lines that are also marked in the spectrum.}
    \label{fig:spec_z}
\end{figure}

An accurate spectroscopic redshift of $z=7.2782 \pm 0.0003$ was derived by obtaining the centroid of the strong \oiii\,$\lambda5007$ emission line seen in the $R\sim1000$ G395M grating spectrum. This redshift was found to be consistent with the weaker \oiii\,$\lambda4959$ as well as the \hb\ line also seen in the grating spectrum. We show the robust detection of the \oiii\,$\lambda5007$ line in the G395M spectrum along with weaker detections of \oiii\,$\lambda4959$ and \hb\ lines, consistent with the spectroscopic redshift, in Figure \ref{fig:spec_z}.

As was noted in \citet{bun23b}, the redshifts derived from emission lines across the $R\sim1000$ gratings as well as the $R\sim 100$ PRISM spectra were found to be consistent, which means that any comparison between the velocity offset of \lya\ emission with respect to the systemic redshift should be reliable and robust.

\subsection{UV magnitude and slope}
As is visible from Figure \ref{fig:R100_spec}, the S/N of the continuum near the \lya\ line in the PRISM spectrum is low. Nonetheless, in this section we attempt to measure the rest-frame UV magnitude at $1500$\,\AA\ as well as the rest-UV slope.

To measure the UV magnitude, we place a boxcar filter centered on $1500$\,\AA\ rest-frame, with a total width of $200$\,\AA\ to boost S/N for magnitude measurement. We then calculate the sigma-clipped median (with clipping applied at $3\sigma$ to calculate the median continuum flux and error in the boxcar filter, which is then used to calculate the UV magnitude and its associated errors. Using this approach, we calculate a UV magnitude of $M_{\rm{UV}} = -17.05$ with $1\sigma$ confidence intervals in the range $(-14.60, -17.74)$, placing JADES-GS-z7-LA at the faint end of the UV luminosity function at $z\sim7-8$ \citep[e.g.,][]{bou22a, don23}.

Although the S/N at rest-frame $\sim1500$\,\AA\ in the PRISM spectrum is not high, we also attempt to infer the rest-frame UV slope ($\beta$) using a Monte Carlo approach. Briefly, the observed flux in the PRISM spectrum in the rest-frame wavelength range $1340-3000$\,\AA\ is perturbed by noise that is randomly drawn from the observed variance of the spectrum. A simple power-law is then fitted to this perturbed spectrum and the best-fitting UV slope is recorded. This process is repeated 500 times, where each time the noise added to the signal is randomly drawn from the variance. The median of these 500 best-fitting power law slopes gives us the UV slope measurement, whereas the standard deviation (which contains the error propagation from the variance) gives us the error on the UV slope measurement. Using this Monte Carlo approach, we measure $\beta = -2.06 \pm 0.14$, which is consistent with what was reported by \citet{cul23} for faint galaxies at similar redshifts. We also find this UV slope measurement to be highly consistent with that measured from SED fitting of NIRCam photometry (details in Section \ref{sec:SED}). The UV slope we measure is consistent with a relatively dust-free environment.

\subsection{Lyman-alpha emission}
\begin{figure}
    \centering
    \includegraphics[width=\linewidth]{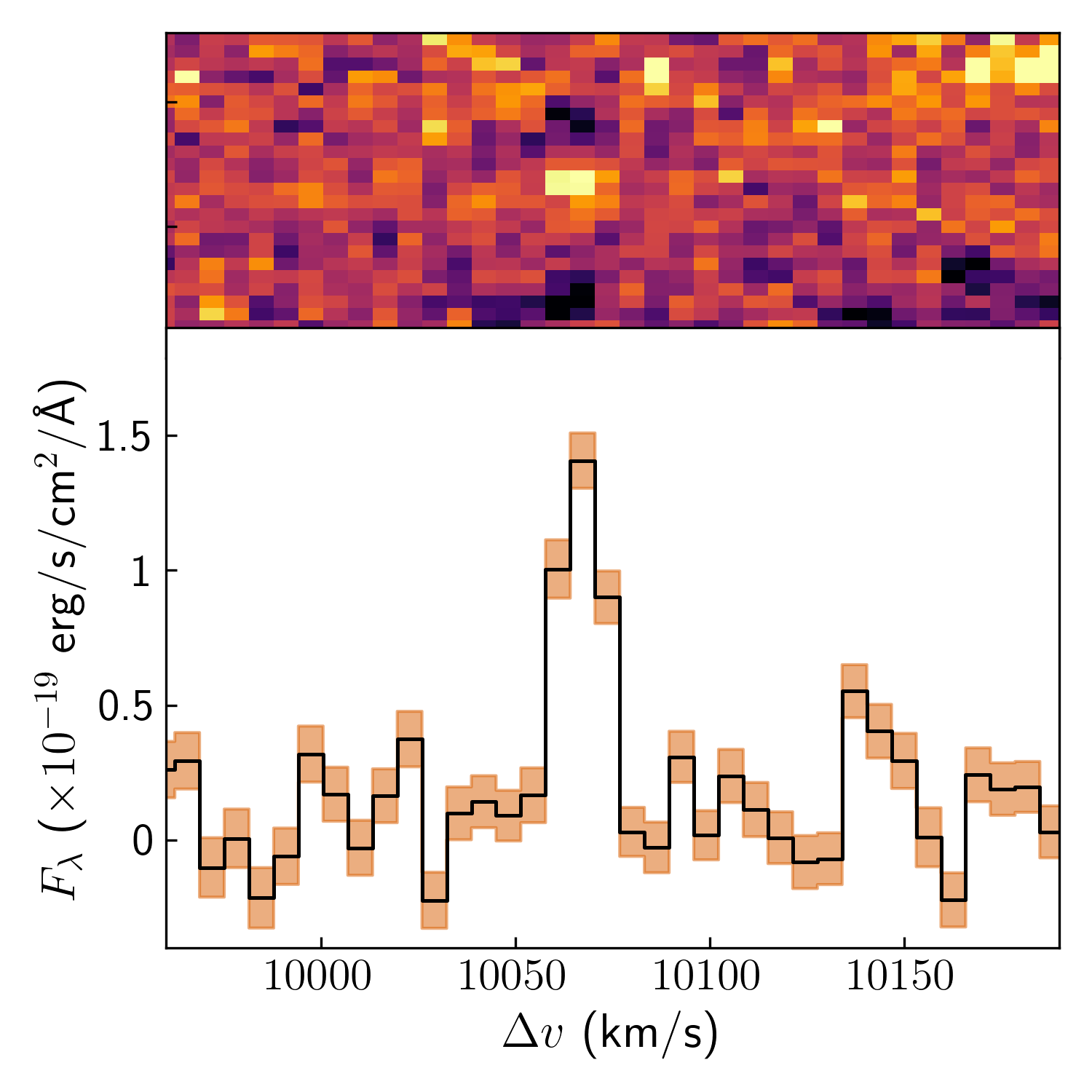}
    \caption{Velocity shift of the \lya\ emission line in the G140M NIRSpec spectrum with a resolution of $R\sim1000$ compared to the systemic redshift ($\Delta v=0$\,\kms) traced by the \oiii\ and \hb\ lines. The emission line appears to be symmetric, centered very close to the systemic redshift with an offset of $113.3\pm80.0$\,\kms, with flux blueward of the peak. The observed line profile is indicative of significant \lya\ photon escape thanks to virtually no neutral gas along the line-of-sight in the galaxy.}
    \label{fig:R1000_spec}
\end{figure}

As discussed in the previous subsection and seen in Figure \ref{fig:R100_spec}, the continuum is not well detected in the PRISM R100 spectrum. However, the \lya\ emission line is clearly detected both in 1D and 2D spectra. The \lya\ line is also clearly detected in the G140M grating spectrum, which corresponds to a spectral resolution of $R\sim1000$ (Figure \ref{fig:R1000_spec}). In this paper, we use the \lya\ line flux measured from the R1000 spectrum, as the spectral resolution in the lower resolution PRISM spectrum is dramatically reduced at shorter wavelengths. 

To measure the emission line properties, we fit a single Gaussian to the emission line in the R1000 spectrum. The continuum around the \lya\ line is not detected in the R1000 spectrum, but is detected with low S/N in the PRISM spectrum. Therefore, we measured the local continuum level from the PRISM spectrum in the wavelength range of rest-frame $1270 - 1380$\,\AA. We note that no N\,\textsc{v} emission is seen in the spectrum and therefore, should not impact the continuum measurement. Owing to the low S/N of the continuum even in the PRISM spectrum, we additionally opted to use a linear extrapolation of fluxes in the F115W, F150W and F200W bands from NIRCam to estimate the continuum around the line, which we found to be highly consistent (within $3\sigma$) with the continuum measured from the PRISM spectrum. 

Given the higher continuum S/N of NIRCam photometry compared to the PRISM spectrum, we opt to use the continuum inferred from extrapolation of NIRCam fluxes. We also note here that given the filter response curve of F115W and the redshift of our source, the \lya\ line lies at the blue edge of the filter, which means that it is likely not contaminating the broad-band filter tremendously, and explains the consistent continuum derived by simple linear extrapolation of NIRCam fluxes and SED fitting. 

Using the single Gaussian fit, we measure a \lya\ line flux of $2.29 \pm 0.25\times 10^{-18}$\,\flux, giving a rest-frame equivalent width of EW$_0 = 388.0 \pm 88.8$\,\AA, making it one of the highest EW \lya\ emitting galaxies currently known in the epoch of reionization \cite[see][for a comparison]{sax23b}. The \lya\ redshift (using a rest-frame wavelength of $1215.67$\,\AA) is found to be $z_{\rm{Ly}\alpha} = 7.2822 \pm 0.0004$.

The \lya\ line has an observed -- line spread function (LSF)-convolved -- full-width at half maximum (FWHM) of $436.7 \pm 56.2$\,\kms. For a point source in the center of the shutter, the LSF at $\sim1.0$ micron is $\approx 208$\,\kms\ (De Graaff et al. in prep), which gives a LSF-deconvolved FWHM of $383.9 \pm 56.2$\,\kms. Perhaps the most striking feature is that the velocity shift of \lya\ emission compared to systemic redshift is low, with $\Delta v = 113.3\pm80.0$\,km/s (where 80\,km/s is the width of the spectral pixel) as shown in Figure \ref{fig:R1000_spec}. We note that this observed \lya\ velocity offset is lower than the LSF at these wavelengths.

\begin{figure}
    \centering
    \includegraphics[width=\linewidth]{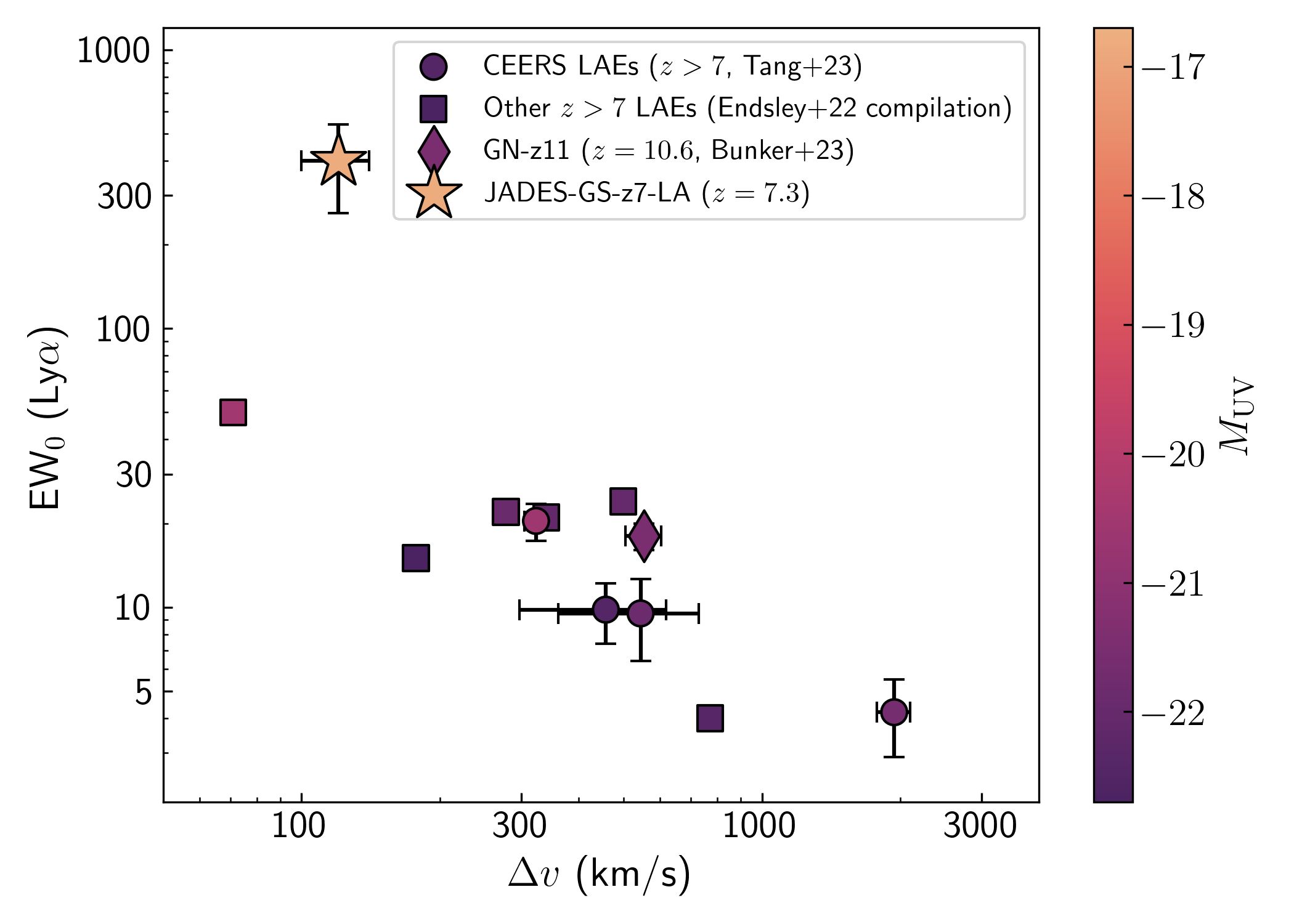}
    \caption{Distribution of \lya\ EW and its velocity offset from the systemic redshift for JADES-GS-z7-LA (star), compared with GN-z11 \citep{bun23} (diamond), $z>7$ LAEs identified from CEERS by \citet{tan23} (circles) and other known $z>7$ LAEs from the \citet{end22c} compilation (squares), which include sources from the ALMA REBELS survey amongst others such as B14-65666 at $z=7.15$ \citep{fur16, has19}, EGS-zs8-1 at $z=7.72$ \citep{oes15, sta17}, COS-zs7-1 at $z=7.15$ \citep{pen16, lap17, sta17} and BDF-3299 at $z=7.11$ \citep{van11, mai15, car17}. We also include the $z=7.68$ galaxy Y002 \citep{val22}. JADES-GS-z7-LA lines up well with the previously observed strong anti-correlation between the strength of the \lya\ emission line and its velocity offset compared to systemic redshift \citep[e.g.,][]{tan23}, and its faint UV magnitude and high \lya\ EW compared to other $z>7$ LAEs stand out.}
    \label{fig:vel_comp}
\end{figure}

We further investigate the implication of high \lya\ EW and low velocity offset in Figure \ref{fig:vel_comp}, where we show the distribution of the \lya\ EW and the observed offset from the systemic redshift for JADES-GS-z7-LA, compared with the recently discovered \lya\ emission from GN-z11 \citep{bun23} and measurements from other $z>7$ galaxies \citep{tan23, end22c}. The points have been color-coded by their absolute UV magnitudes. The distribution is in line with an observed anti-correlation between \lya\ EW and velocity offset from the systemic redshift, whereby UV-brighter galaxies also tend to show larger velocity offsets \citep[see][for example]{tan23}. Figure \ref{fig:vel_comp} also demonstrates the extremely high \lya\ EW as well as faint continuum of JADES-GS-z7-LA compared to other known LAEs at $z>7$. 

Such high \lya\ line strengths and narrow \lya\ profiles peaking close to the systemic redshift are often indicative of the presence of low column density channels in the neutral gas surrounding H\,\textsc{ii} regions, or very little neutral gas, which enables the escape of \lya\ photons without considerable absorption and scattering out of the line-of-sight \citep[see][for example]{ver17}. We discuss this further in Section \ref{sec:reionization}. 

\subsection{Other strong emission lines}
In the R100 PRISM spectrum, \oii\,$\lambda3727$, \hb\ and \oiii\,$\lambda\lambda 4959, 5007$\,\AA\ lines are clearly detected. Since no continuum is detected near these rest-optical emission lines, the continuum level is estimated by using linear extrapolations from NIRCam photometry corresponding to the bands closest to the lines. By fitting single Gaussian functions, we measure rest-frame equivalent widths of EW$_0$(\oii\,$\lambda3727$) $=69.9 \pm 14.4$\,\AA, EW$_0$(\hb) $= 113.0 \pm 23.1$\,\AA, EW$_0$(\oiii\,$\lambda 4959$) $=330.3 \pm 62.9$\,\AA\ and EW$_0$(\oiii\,$\lambda 5007$) $=854.8 \pm 58.0$\,\AA, which are typical of what is observed in $z\sim7$ galaxies \citep[e.g.,][]{end21, end22c}. 

As previously noted, the \oiii\,$\lambda 5007$line is also clearly detected in the G395M spectrum. Using the same continuum level as for the PRISM line measurements, the \oiii\,$\lambda5007$ line flux measured from the G395M spectrum is $7.2\pm0.6 \times 10^{-19}$\,\flux, with EW$_0 = 1028.1 \pm 165.5$\,\AA. The \hb\ line is weaker, but it is possible to fit a single Gaussian to the line, which gives a line flux of $1.4 \pm 0.6 \times 10^{-19}$\,\flux, with EW$_0 = 204.9 \pm 62.5$\,\AA. Since the higher resolution grating is more sensitive to narrow line emission, the flux measured from the grating is higher than that measured from the lower resolution PRISM spectrum. The \oiii\,$\lambda5007$ line appears to be narrow, with FWHM $=254.8 \pm 23.7$\,\kms\ and no obvious broad wings. The emission line measurements are given in Table~\ref{tab:measurements}.

Using these lines we measure rest-optical line ratios that trace the ionization state as well as metal content of the ISM. For consistency, we calculate the line ratios by using all the line fluxes measured from the PRISM spectrum. We find a \oiii$\,\lambda 5007$/\oii\,$\lambda3727$ ratio (O32) of $11.1 \pm 2.2$, (\oiii\,$\lambda \lambda 4959,5007$ + \oii\,$\lambda3727$/\hb) (R23) ratio of $11.2 \pm 2.6$ and \oiii\,$\lambda 5007$/\hb\ (R3) ratio of $7.6 \pm 1.5$, which are all consistent with low metallicities and high ionization parameters in the ISM \citep{tru23, cur23, kat23b, san23, cam23, nak23}. However, we find that the O32 ratio we measure is on the lower end of the range of ratios reported in much brighter LAEs at $z>7$ by \citet{tan23}, although a few of those were derived from photometry alone. The line ratios are also summarized in Table \ref{tab:measurements}.

\begin{table}[t]
    \centering
    \caption{Observed spectroscopic properties of JADES-GS+53.16746-27.7720 (or JADES-GS-z7-LA).}
    \begin{tabular}{l r}
    \toprule 
    Parameter & Measurement  \\
    \midrule 
    $z_{\rm{sys}}$ & $7.2782 \pm 0.0003$ \\
    $z_{\rm{Ly}\alpha}$ & $7.2822 \pm 0.0004$ \\ \\
    \textit{R100 line fluxes [$\times 10^{-20}$\,\flux]} \\
    F(\oii\,$\lambda 3727$)  & $5.4\pm 1.4$ \\
    F(\hb)  & $7.9\pm 2.2$ \\
    F(\oiii\,$\lambda 4959$)  & $23.1 \pm 6.0$ \\
    F(\oiii\,$\lambda 5007$)  & $59.9 \pm 4.0$ \\ \\

    \textit{R100 equivalent widths [\AA]} \\
    EW$_0$(\oii\,$\lambda 3727$) & $69.9 \pm 14.4$ \\   
    EW$_0$(\hb) & $113.0 \pm 23.1$ \\   
    EW$_0$(\oiii\,$\lambda 4959$) & $330.3 \pm 62.9$ \\ 
    EW$_0$(\oiii\,$\lambda 5007$) & $854.8 \pm 58.0$ \\ \\

    \textit{R1000 line fluxes [$\times10^{-20}$\,\flux]} \\
    F(\lya)  & $229.2 \pm 25.9$ \\
    F(\hb)  &   $14.3 \pm 5.9$ \\
    F(\oiii\,$\lambda 5007$)  & $64.6 \pm 6.2$ \\ \\

    \textit{R1000 equivalent widths [\AA]} \\
    EW$_0$(\lya) & $388.0 \pm 88.8$ \\
    EW$_0$(\hb) &   $204.9 \pm 62.5$ \\
    EW$_0$(\oiii\,$\lambda 5007$) & $1028.1 \pm 165.5$ \\ \\

    \textit{Line ratios (from PRISM)} \\
    \oiii\,$\lambda 5007$/\oii\,$\lambda 3727$ (O32) & $11.1 \pm 2.2$\\
    (\oiii\,$\lambda \lambda 4959,5007$ + \oii)/\hb\ (R23) & $11.2 \pm 2.6$ \\
    \oiii\,$\lambda 5007$/\hb\ (R3) & $7.6 \pm 1.5$ \\ 
    \hg/\hb\ & $<0.49$ \\

    \bottomrule
    \end{tabular}
    \label{tab:measurements}
\end{table}

\subsection{Balmer lines, dust and UV slope}
Although the \hb\ line is clearly detected in the spectrum, the \ha\ line is redshifted out of the observed wavelengths in the R100 spectrum and \hg\ is not detected. We derive a $1\sigma$ upper limit of \hg/\hb$<0.49$, which is consistent with expectations from Case B recombination for an electron density, $n_e = 100$\,cm$^{-3}$ and temperatures of up to 20,000\,K \citep{dop03, ost06}. However, we note that Case B recombination may not be the most appropriate approximation for galaxies that may be leaking LyC photons, which we expand upon further in Section \ref{sec:lya_escape}.

This limit, unfortunately, is not useful to place constraints on the dust attenuation in the galaxy spectrum. However, given the extremely high equivalent width of the \lya\ line combined with the narrow profile peaking close to the systemic redshift, one might expect there to be little to no dust in the system, which enables \lya\ photons to escape along the line of sight evading dust absorption.

We note that any correction due to dust in the neutral ISM will boost the \lya\ flux. However, it has been shown that due to the complex radiative transfer involved in the propagation of \lya\ photons and the geometry of the ISM, dust may not play a huge role in \lya\ escape \citep[e.g.,][]{ate08, fin10}. Owing to a lack of robust spectroscopic indicators of dust attenuation as well as the strong \lya\ emission seen from this galaxy, we assume dust-free conditions going forward.

\subsection{Photometric measurements and the presence of a companion source} 
\label{sec:photometry}
We report the measured fluxes and errors from the NIRCam images in Table \ref{tab:nircam_mags}. Due to the presence of strong \oiii\ and \hb\ emission, JADES-GS-z7-LA is brightest in the F410M medium band and the flux in the F115W band appears to be boosted due to \lya, even though \lya\ is close to the blue edge of the filter transmission profile.

\begin{figure*}
    \centering
    \includegraphics[width=\textwidth]{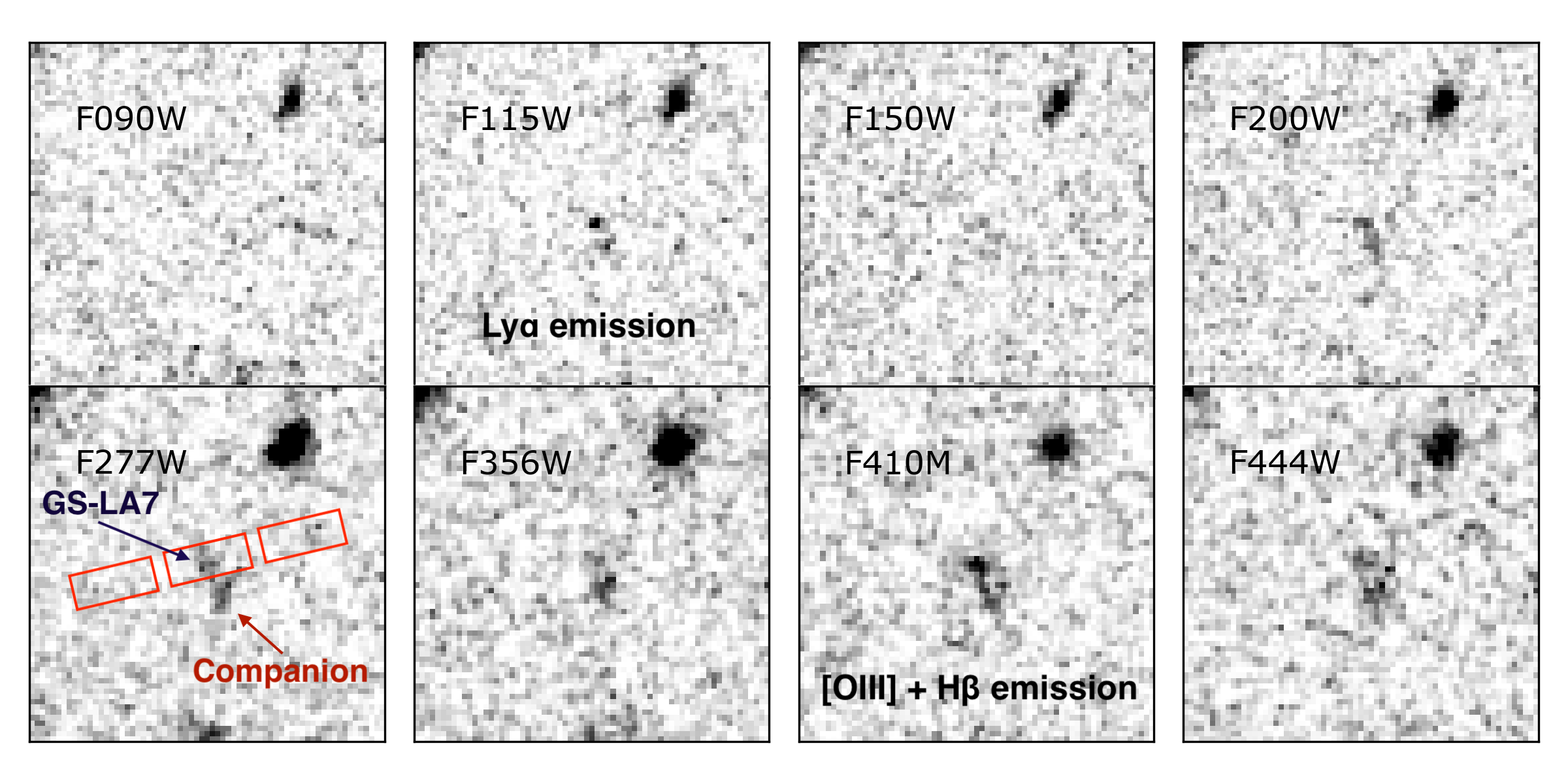}
    \caption{NIRCam cutouts ($1.5''\times1.5''$) showing the location of the \lya\ emitting galaxy and the companion source. Also shown in red in the bottom-left panel is the location of the NIRSpec MSA shutters that led to the discovery of \lya\ emission from JADES-GS-z7-LA, but did not include the companion source. Both galaxies drop out in the F090W filter and qualitatively show similar colors across the broadband filters, suggesting that the companion source likely has a photometric redshift comparable to JADES-GS-z7-LA. The F115W filter is contaminated by \lya\ emission at $z\sim7.3$, which explains the relatively bright flux from JADES-GS-z7-LA and may also explain emission from the companion if it is at the same redshift.}
    \label{fig:nircam}
\end{figure*}

We also note that there is a companion source next to JADES-GS-z7-LA visible in the NIRCam imaging, separated on the sky by $0.23''$. This source unfortunately fell outside the MSA shutter targeting JADES-GS-z7-LA. To our knowledge, this source has not appeared in any previous \emph{HST} selected catalogues. The presence of this companion meant that careful de-blending of the two sources in the NIRCam images was needed to isolate the fluxes of each of these two galaxies. The two sources were de-blended using the {\sc deblend\_sources} function of \textsc{photutils} with a detection threshold of $2\sigma$ and deblending parameters nlevels~$=32$ and contrast~$=0.001$. Aperture photometry was then performed centered on the respective centroid of the de-blended sources.

From Figure \ref{fig:nircam}, it is clear that the companion source also drops out in F090W, with potential contribution from \lya\ emission in F115W (although the line falls close to the edge of the transmission) and appears to be bright in F410M (F356W$-$F410M $= 0.81$), which may be explained by contribution of \oiii\ and \hb\ if the companion source also lies at a similar redshift to that of JADES-GS-z7-LA. We note that both sources appear to be fainter than expected in the F150W band, which we attribute to issues with local background subtraction. 

The companion, if considered to be at the same redshift, has a fainter absolute UV magnitude at $1500$\,\AA, $M_{\rm{1500}} = -16.5$ and a shallower UV slope, $\beta = -1.8 \pm 0.2$. To accurately infer the physical properties of both the LAE and the companion source, we employ spectral energy distribution (SED) fitting using fluxes measured from aperture photometry, which we describe in Section \ref{sec:SED}.
\begin{table}[]
    \centering
    \caption{\emph{JWST}/NIRCam background subtracted fluxes and errors (in nanoJanskys) of JADES-GS-z7-LA and the companion source.}
    \begin{tabular}{l c c}
    \toprule
    Filter  &   JADES-GS-z7-LA  &   Companion   \\
      &   (nJy)  &   (nJy)  \\
    \midrule
    F090W   &   $0.34\pm0.90$   &   $1.00\pm0.91$ \\
    F115W   &   $3.17\pm0.63$ &   $3.22\pm0.63$ \\   
    F150W$^*$   &   $1.43\pm0.76$ &   $2.22\pm0.76$ \\
    F200W   &   $2.99\pm0.64$ &   $3.76\pm0.64$ \\    
    F277W   &   $3.33\pm0.31$ &   $3.37\pm0.31$ \\
    F356W   &   $2.44\pm0.31$ &   $4.24\pm0.31$ \\
    F410M   &   $11.52\pm0.60$ &   $8.99\pm0.59$ \\
    F444W   &   $7.28\pm0.56$ &   $7.52\pm0.56$ \\
    \bottomrule
    \end{tabular}

    $^*$ The F150W image suffers from local background subtraction issues, making the measured fluxes lower than expected.
    \label{tab:nircam_mags}
\end{table}

\subsection{star formation rate derived from Balmer emission}
We can estimate the \ha\ line flux using the observed \hb\ line flux in a Case B recombination scenario. Assuming no dust, $n_e = 100$\,cm$^{-3}$ and $T_e = 15,000$\,K, which is typical of high redshift galaxies (e.g. \citealt{cur23}) and adopting an intrinsic \ha/\hb\ $\approx2.788$ from \textsc{pyneb} \citep{pyneb}. As noted previously, the \hb\ flux is higher in the G395M grating spectrum compared to the PRISM spectrum. Therefore, we estimate an \ha\ line flux of $2.9\times10^{-19}$\,\flux\ using the \hb\ flux from PRISM corresponding to \ha\ luminosity of $1.4\times 10^{41}$\,erg\,s$^{-1}$, and \ha\ flux of $4.0 \times 10^{-19}$\,\flux\ and luminosity of $2.6\times10^{41}$\,erg\,s$^{-1}$ from the grating spectrum.

It has been shown that lower metallicity galaxies at high redshifts are capable of achieving high ionizing photon production efficiencies at the same star formation rates when compared to lower redshift, more chemically evolved galaxies \citep[e.g.,][]{cha01}. Therefore, using the conversion factor derived from stellar population synthesis models with $\sim$5\% solar metallicity, by \citet{red22}, we multiply the \ha\ luminosity with $2.12\times10^{-42}$\,M$_\odot$\,yr$^{-1}$\,erg\,s$^{-1}$, giving SFR(\ha) $=0.29-0.56$\,M$_\odot$\,yr$^{-1}$. This SFR is lower when compared to the standard \citet{ken98} relation, which assumes a \citet{sal55} IMF and gives SFR $=0.75 - 1.43$\,M$_\odot$\,yr$^{-1}$. 

The measured SFR(\ha) of $\approx0.3-0.6$\,M$_\odot$\,yr$^{-1}$, which given the extremely faint $M_{\rm1500} \approx -17.0$ places the galaxy comfortably above the star-forming main sequence at $z>6$ \citep{sha23}, consistent with expectations from a galaxy with strong emission lines including \lya.

\subsection{Ionizing photon production efficiency}
Using the observed Balmer line emission and the monochromatic UV luminosity at $1500$\,\AA, we can also infer the ionizing photon production efficiency ($\xi_{\rm{ion}}$). The number of ionizing photons produced is closely linked to the (dust-corrected) \ha\ luminosity. Assuming Case B recombination and under the same assumptions of the electron density and temperature as in the above section, this relation is given by $N(H^0) =\,$L(H$\mathrm{\alpha}$)\,$\times\,7.37 \times 10^{11}$ \citep{ost06}. The $\xi_{\rm{ion}}$ is then calculated as $\xi_{\rm{ion}} = N(H^0)/L_{\rm{1500}}$.

Using the $L_{\rm{1500}}$ measured directly from the PRISM spectrum and estimating the \ha\ line luminosity using \hb\ with the assumptions outlined in the previous subsection, we infer a $\log(\xi_{\rm{ion}}/\rm{erg\,Hz}^{-1}) = 25.66 \pm 0.14$ when using the \hb\ flux measured from the PRISM spectrum, and $\log(\xi_{\rm{ion}}/\rm{erg\,Hz}^{-1}) = 25.82 \pm 0.15$ when using the \hb\ flux from the medium resolution grating spectrum, which is theoretically more sensitive to line emission. This \xiion\ value is higher than the canonical values of $25.2-25.3$ assumed by simple reionization models \citep[e.g.,][]{rob13}, and consistent with what is typically seen for LAEs at $z>6$ \citep[e.g.,][]{tan23, sim23, sax23b}. 

In the following section we fit SED models to both JADES-GS-z7-LA and the companion source and derive other physical properties of these sources.

\section{Insights from spectral energy distribution fitting}
\label{sec:SED}

Fitting SED models to spectra, photometry, or both simultaneously is crucial to gain valuable insights about the nature of distant galaxies. Our aim is to find best-fitting SEDs for both JADES-GS-z7-LA and its close companion to determine whether the redshift of the companion source is consistent with that of JADES-GS-z7-LA, and then infer their physical properties. We start our analysis by simultaneously fitting the spectrum and photometry of JADES-GS-z7-LA to derive accurate constraints on the star formation history as well as the ISM properties of this galaxy. Such spectro-photometric fitting can lead to more robust estimates on the posterior distributions of key galaxy physical properties, primarily driven by constraints provided by the emission lines.

We use the \textsc{python} code \textsc{bagpipes}, which stands for Bayesian Analysis of Galaxies for Physical Inference and Parameter Estimation \citep{car18} to perform SED fitting. We use templates that include both stellar emission, based on \citet{bc03} stellar population synthesis models constructed using the \citet{kro01} initial mass function (IMF) with initial masses in the range $0.6-120$\,M$_\odot$, as well as nebular continuum and line emission. The nebular emission is computed using the photoionization code Cloudy \citep{fer13}. We note here that the metallicity of the line-emitting gas and stars is assumed to be the same.

We use a two-component star formation history: the first is an exponentially declining ($\tau$-model) star formation history, and the second is a recent burst of star formation. For the exponentially declining model, we let the age of the stars formed vary from 0.5~Myr to the age of the Universe at the redshift of the source, let the $\tau$ vary from 300 Myr to 10 Gyr, with a total stellar mass formed in the range $10^6 - 10^{11}$\,$M_\odot$ with permitted stellar metallicities in the range $0.01-1.2$\,$Z_\odot$. 

For the burst component, we allow the age of the burst to vary between $1-20$\,Myr and the mass formed during the burst to vary between $10^6-10^{11}$\,$M_\odot$. We include the dust attenuation following the \citet{sal18} law, which is very close to the SMC extinction curve \citep{gor03} and is a good approximation for high redshift galaxies \citep[e.g.,][]{shi20}. For nebular emission, we allow the dimensionless ionization parameter, $\log(U)$, to vary between $-0.5$ to $-3$.
\begin{figure*}
    \centering
    \includegraphics[width=\textwidth]{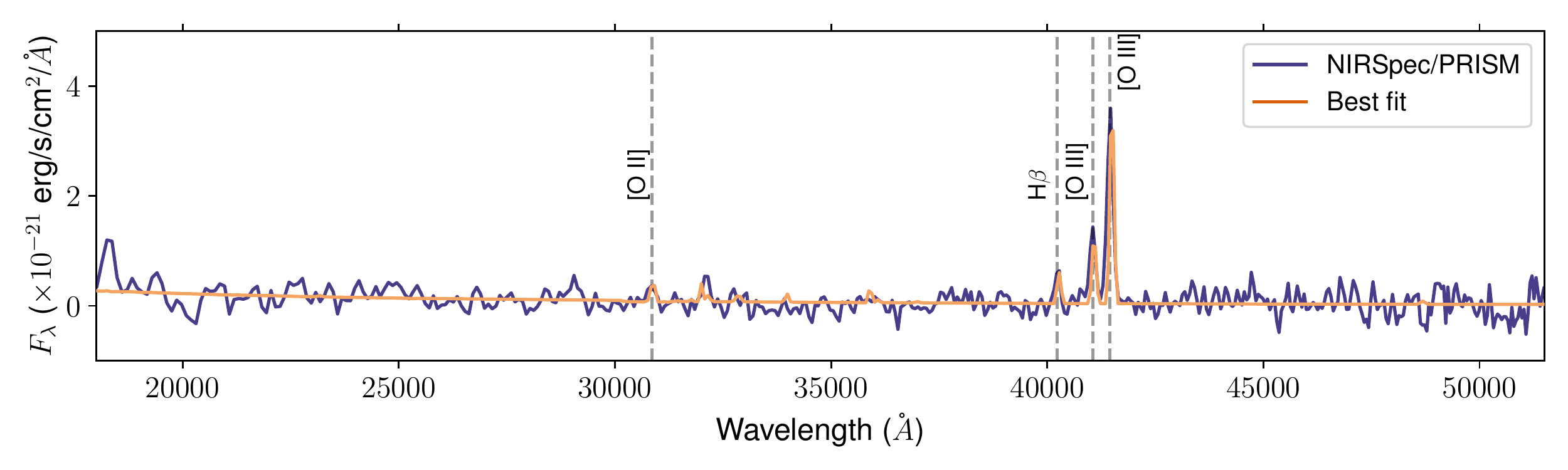}
    \caption{NIRSpec/PRISM spectrum of JADES-GS-z7-LA (excluding the \lya\ region) with the best-fitting spectrum from \textsc{bagpipes}. We fit the entire spectrum, but only highlight the strongest emission lines in the figure that have the most constraining power. The fits to the observed emission lines suggest that a high ionization parameter, $\log(U) = -0.72 \pm 0.14$ and low metallicities of $Z_{\rm{burst}} = 0.31 \pm 0.26\,Z_\odot$ are needed to explain the observed emission lines. The best-fit SED also favors a bursty star formation history (see text). The constraints derived from fitting the spectrum are useful to then fit the photometry of the galaxy as well, as shown in Figure \ref{fig:SED}.}
    \label{fig:spec_fit}
\end{figure*}

We note here that under normal circumstances the increasingly neutral IGM beyond redshift 6 would suppress nearly all \lya\ emission blueward of the peak. However, we know from spectroscopy that JADES-GS-z7-LA is a strong \lya\ emitter, suggesting that it likely is situated in an ionized bubble within a generally neutral Universe at $z\approx7.3$. Additionally, the complex radiative transfer calculations required to accurately model \lya\ emission are not currently included in \textsc{bagpipes}. Therefore, when fitting the observed spectrum, we mask the region containing \lya. Additionally, when fitting the photometry, we turn off the IGM attenuation, mimicking conditions inside an ionized bubble with little to no neutral hydrogen. This results in the best-fitting galaxy physical parameters predicting a \lya\ flux that can then be matched to the observed F115W photometry.

The PRISM spectrum and NIRCam photometry of JADES-GS-z7-LA are best-fit with a ``bursty'' star formation history, with $\approx53.2\%$ of the total current stellar mass of $10^{6.9}$\,M$_\odot$ (with 16th and 84th percentile confidence interval [$10^{6.8}$, $10^{7.1}$\,M$_\odot$]) with a recent burst with age$_{\rm{burst}} = 2.2\,([1.3, 10.3])$\,Myr and metallicity $Z_{\rm{burst}} = 0.39\,([0.11, 0.70])$\,$Z_\odot$. The median age of stars formed via exponentially declining history is higher, with age$_{\rm{exp}} = 8.8\,([2.0, 61.9]$)\,Myr with a broad range, and metallicity $Z_{\rm{exp}} = 0.29\,([0.06, 0.68])$\,$Z_\odot$. 

As expected, a high $\log(U)$ of $-0.94\,([-1.24, -0.67])$ is needed to explain the observed emission lines in the spectrum. The best-fitting SED also requires a small dust attenuation of $A_V = 0.24\,([0.17, 0.30])$\,mag, which may be incompatible with the observed strong \lya\ emission seen in the spectrum. The best-fitting SED to the PRISM/CLEAR spectrum of JADES-GS-z7-LA is shown in Figure \ref{fig:spec_fit}.
\begin{figure}
    \centering
    \includegraphics[width=\linewidth]{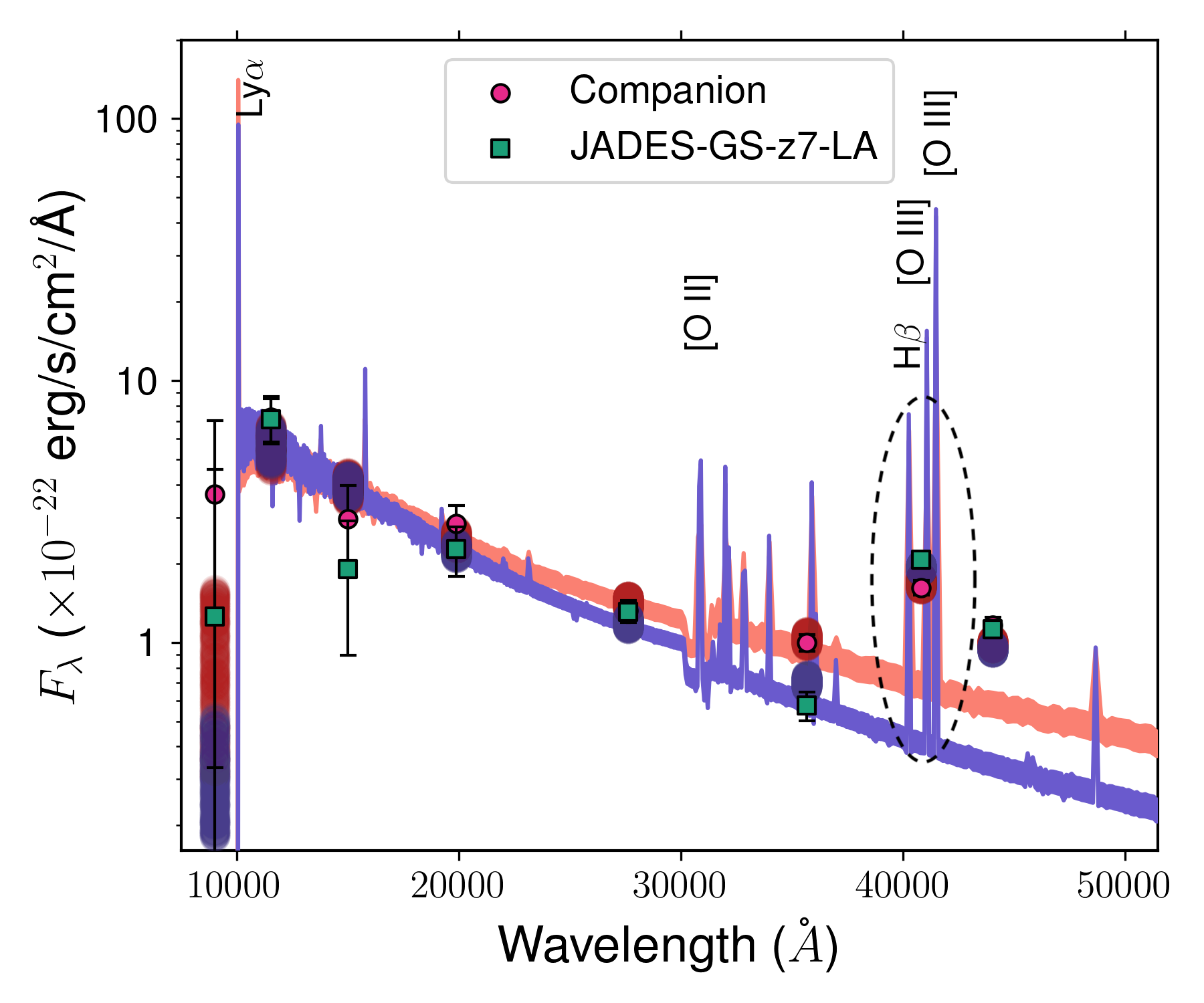}
    \caption{Best-fitting SEDs of JADES-GS-z7-LA and its companion source, fixing the redshift to be at $z=7.278$ for both objects. We find that a good fit is obtained for the companion source at the same redshift as JADES-GS-z7-LA, which in particular can explain the photometric excess observed in the F410M band (dashed oval). Overall, JADES-GS-z7-LA has a `bluer' SED, which is reflected in the preference for a bursty star formation history when compared to the companion source.} 
    \label{fig:SED}
\end{figure}

Under the assumption that the companion source is at the same redshift as JADES-GS-z7-LA, we use the same set of priors to find the best-fitting SED for the companion. Since there is no spectrum available, we only use the NIRCam photometry for the fitting. For the companion, we find that the fraction of stellar mass formed in a recent burst of $\approx5$\,Myr is only $\approx 13.8\%$ of the total stellar mass of $10^{7.2} \,([10^{7.0}, 10^{7.5}])$\,M$_\odot$, suggesting a less ``bursty'' star formation history compared to JADES-GS-z7-LA. The median age of stars in the burst is age$_{\rm{burst}} = 3.5\,([1.5, 12.3])$\,Myr, with metallicity $Z_{\rm{burst}} = 0.13\,([0.03, 0.55])$\,$Z_\odot$. The age of stars formed following exponential decline is age$_{\rm{exp}} = 33.6\,([8.0, 111.0])$\,Myr, with metallicity $Z_{\rm{exp}} = 0.16\,([0.04, 0.29])\,Z_\odot$.

The best-fitting SED prefers a relatively lower ionization parameter of $\log(U) = -1.29\,([-1.85, -0.81])$, but it is difficult to place accurate constraints on this without spectroscopy and requires slightly more dust attenuation, with $A_V = 0.28\,([0.23, 0.35])$\,mag compared to the LAE. The best-fitting SEDs of both JADES-GS-z7-LA and the companion source are shown in Figure \ref{fig:SED} with a summary of their inferred physical properties given in Table \ref{tab:sed_fits}.

\begin{table}
    \centering
    \caption{Physical properties derived from SED fitting for JADES-GS-z7-LA and the companion source, assuming an exponentially declining star formation history with a recent burst of star formation.}
    
    \begin{tabular}{l c c}
    \toprule
    Property & JADES-GS-z7-LA & Companion \\
    \midrule 
    $M_{\rm{UV}}$ & $-17.0 \pm 1.0$ & $-16.9 \pm 1.0$ \\
    \textbeta\ slope & $-2.0 \pm 0.2$ & $-1.7 \pm 0.2$ \\
    $\log(M_\star^{\rm{total}}/M_\odot)$ & $6.9^{+0.2}_{-0.1}$ & $7.2^{+0.3}_{-0.2}$ \\
    $M_\star^{\rm{burst}}/M_\star^{\rm{total}}$ & $53.2\%$ & $13.8\%$ \\
    age$_{\rm{burst}}$ [Myr] & $2.2^{+6.9}_{-0.9}$ & $3.1^{+5.6}_{-1.6}$ \\ 
    $Z_{\rm{burst}}$ [$Z_\odot$] & $0.31^{+0.40}_{-0.26}$ & $0.13^{+0.44}_{-0.09}$ \\
    age$_{\rm{exp}}$ [Myr] & $4.1^{+45.9}_{-2.1}$ & $32.9^{+57.1}_{-23.1}$ \\
    $Z_{\rm{exp}}$ [$Z_\odot$] & $0.29^{+0.30}_{-0.21}$ & $0.12^{+0.16}_{-0.09}$ \\
    $\log(U)$ & $-0.94^{+0.15}_{-0.13}$ & $-1.29^{+0.49}_{-0.72}$ \\    
    \bottomrule    
    \end{tabular}
    \label{tab:sed_fits}
\end{table}

Based on SED fitting, and particularly the photometric excess in the F410M band as well as possible emission line contribution in the F115W band due to \lya, it is likely that the companion source is at the same redshift as that of JADES-GS-z7-LA. It may also be possible that these two individual sources represent star-forming regions of the same larger system. It is also possible that the two sources are merging, but there is no spectroscopic evidence for this currently. The SED of JADES-GS-z7-LA clearly suggests that it is a bursty, actively forming galaxy whereas the companion seems to have formed the bulk of its stars earlier than JADES-GS-z7-LA.

In light of the high EW \lya\ emission seen from JADES-GS-z7-LA, questions naturally arise about the contribution of these two nearby sources in creating a bubble of ionized gas surrounding them. Any enhancement in the escape of ionizing photons due to association/interaction may have implications on the mechanisms via which \lya\ and potentially LyC ionizing photons can escape from galaxies in the reionization era. In the following section, we explore the spectroscopic indicators of \lya\ and LyC escape from JADES-GS-z7-LA, and investigate whether the LAE and its companion would be capable of inflating an ionized bubble large enough to explain the observed \lya\ emission.

\section{Implications for ionizing photon escape and reionization}
\label{sec:reionization}

In this section we explore the implications of the observations presented in this paper on the state of the IGM surrounding JADES-GS-z7-LA and its companion. The clear presence of \lya\ emission from JADES-GS-z7-LA at $z\approx7.3$ offers a unique opportunity to investigate the IGM conditions within its vicinity as well as the production and escape of ionizing photons in an LAE at $z>6$, and we begin our analysis by first constraining the escape fraction of \lya\ photons from this galaxy.

\subsection{Lyman-alpha escape fraction}
\label{sec:lya_escape}
Using Balmer emission lines, it is possible to derive an escape fraction for \lya\ under certain assumptions. For example, assuming Case B recombination, that is no LyC photon escape from the galaxy, and a dust free environment, $n_e = 100$\,cm$^{-3}$ and $T_e = 15,000$\,K, the intrinsic \lya/\hb\ ratio is $23.3$ \citep{ost89}. The \fesc(\lya) can then be calculated as: \fesc(\lya) = L(\lya)/($23.3\,\times\,$L(\hb)). 

As was previously noted, the \hb\ line flux measured from the G395M grating spectrum is higher than that measured from PRISM. The \hb\ line flux from the PRISM spectrum results in \fesc(\lya) $= 1.24 \pm 0.25$, whereas that from the grating results in \fesc(\lya) $= 0.68 \pm 0.20$. Therefore, we find that under Case B recombination the \lya\ escape fraction for JADES-GS-z7-LA is realistically in the range \fesc(\lya) $\approx 0.7 - 1.0$, which is consistent with expectations from LAEs in a dust-free environment \citep[e.g.,][]{dij16}.

A very high ($\sim100\%$) \lya\ escape fraction was also reported for a LyC leaking galaxy in the local Universe by \citet{izo18a}, which also had a LyC escape fraction of \fesc(LyC) $\approx0.46$. This implies that Case B recombination may not necessarily be valid for galaxies that may be leaking a significant amount of LyC photons through an optically thin ISM, where Case A recombination is more appropriate. In Case A, it is possible to achieve $\sim1.5\times$ higher \lya/\hb\ ratios, due to the effective recombination coefficient of Balmer series transitions being $\approx1.5$ times higher compared to Case B. Using a $1.5\times$ intrinsic ratio following Case A, we would infer \fesc(\lya) $\approx 0.45 - 0.83$.

We also note that the intrinsic \lya/\hb\ ratio is more sensitive to the assumed electron density than temperature. For example, assuming $n_e = 5000$\,cm$^{-3}$ (using {\sc pyneb}) can increase the intrinsic ratio to $\approx30$, leading to even lower \fesc(\lya). We also note that the inclusion of any dust correction in this calculation will actually increase the observed \lya/\hb\ ratio, leading to even higher \fesc(\lya).

Several studies in the literature, both using theory and observations, have attempted to link the leakage of LyC photons to \lya\ escape \citep[e.g.,][]{mar18, gaz20, izo21}, \lya\ luminosities \citep[e.g.,][]{maj22} and \lya\ line profiles \citep[e.g.,][]{ver17, nai22}. Based on the high \fesc(\lya) and its line profile, it is not unreasonable to expect that JADES-GS-z7-LA is leaking a significant amount of LyC radiation. In the following section, we investigate indications of LyC leakage using other spectroscopic and photometric indicators.

\subsection{Lyman continuum leakage from JADES-GS-z7-LA}
Although \emph{JWST} imaging and spectroscopy has now made it relatively easy to measure the distribution of UV luminosities and \xiion\ of galaxies in the reionization era, placing accurate constraints on \fesc\ remains a big challenge as direct measurements of LyC emission are extremely unlikely at $z>4$ \citep{ino14}. The shape and the strength of \lya\ emission from galaxies can help trace Lyman continuum leakage -- in particular, a double peaked \lya\ profile with transmission of flux blueward of the systemic redshift can trace physical conditions representative of low column density of neutral gas in the ISM, which is also conducive to LyC photon leakage \citep{ver15}. After processing through the IGM, a resulting small shift ($v \leq 150$\,\kms) in the peak of a single \lya\ emission line compared to the systemic redshift can then be indicative of the presence of an ionized region surrounding the galaxy, likely driven by significant LyC leakage \citep[e.g.,][]{mas20}.

There are several other spectroscopic indicators of LyC \fesc\ proposed in the literature, including high O32 ratios \citep[e.g.,][]{fai16}, low covering fractions of low ionization interstellar absorption features (\citealt{hec01}; see e.g. \citet{sal23} and references therein for more recent works) and the presence of high ionization emission lines \citep[e.g.,][]{sch22a, sax22b}. Steep rest-frame UV slopes as well as reduced strengths of rest-frame nebular emission lines have also been suggested as tracers of significant LyC leakage \citep[e.g.,][]{top22}. 

JADES-GS-z7-LA is one of the highest EW LAEs currently known in the reionization epoch with rest-frame EW of $=388.0 \pm 88.8$\,\AA, and a narrow (unresolved) emission line profile, which is only offset from its systemic redshift by $113.3$\,\kms\ (Figure \ref{fig:R1000_spec}) suggesting the presence of channels of ionized gas through which LyC photons could readily escape. Based on the empirical correlations between \lya\ properties and LyC leakage derived by \citet{ver17}, we note that having EW(\lya) $=388$\,\AA\ with \fesc(\lya) $\approx 0.7-1.0$ would imply \fesc(LyC) $> 0.05$ (see Figure 2 of \citealt{ver17}, see also \citealt{maj22}). We do note here, however, that there is considerable scatter in the relations between \lya\ properties and LyC escape, in part driven by the resonant nature of the \lya\ line which readily undergoes scattering and can be easily removed from the line of sight \citep[see][for example]{dij16}, possibly making \lya\ an unreliable probe of LyC leakage.

Other diagnositics that are not as line of sight dependent as \lya\ may be more reliable to infer LyC escape fractions \citep[e.g.,][]{cho23}. Significant LyC escape may also be inferred from the high EW$_0$(\oiii+\hb) and O32 measured from the spectrum of JADES-GS-z7-LA \citep[see][and references therein]{sax22a}, where the mild correlation observed between \fesc(LyC) and the strength of \oiii+\hb\ emission would suggest \fesc(LyC) $>0.2$. However, the relationship between these two quantities also has considerable scatter, and studies have shown that high \oiii+\hb\ line strengths may also be achieved without significant LyC leakage \citep[e.g.,][]{paa18}.

\citet{zac13} also showed that the escape of LyC photons either from ionization-bounded \hii\ regions with holes blown across them by star formation and/or supernovae, or a density-bounded \hii\ region photo-ionized by sources of hard radiation fields (e.g. low metallicity young stars) can be probed by using a combination of the UV slope and strength of the \hb\ line. \citet{zac13} showed that the strength of the \hb\ EW actually decreases with increasing \fesc(LyC). The \hb\ EW can be used in combination with the steepness of the UV slope to infer \fesc(LyC), where UV slopes of $\beta < -2.4$ required for most assumed star formation histories. 

JADES-GS-z7-LA shows a strong \hb\ emission (EW$_0 \approx 113-205$\,\AA), but does not show a particularly steep UV slope ($\beta \approx -2.0$). Using the parameter space probed by the models of \citet{zac13}, no significant \fesc(LyC) can be inferred from a galaxy with $\beta = -2.0$ and EW$_0$(\hb) $\approx150$ \citep[see also][]{chi22}. \citet{top22} recently showed that even the inclusion of dust requires the observed UV slopes to be steeper than $-2.4$ to be able to infer LyC leakage from a system. A redder UV slope likely traces the presence of nebular continuum emission, meaning that a large fraction of LyC photons are being reprocessed within the galaxy to produce nebular continuum and line emission. The presence of dust can also redden the UV slope, which is also detrimental to LyC leakage. Although, the detection of strong \lya\ emission can help rule out the presence of dust (at least along the line of sight), the presence of strong nebular emission lines points towards nebular contribution being the likely reason for a relatively shallow UV slope that is observed. 

By exploring multiple indicators of LyC leakage, we find that although certain spectroscopic indicators, most importantly the \lya\ escape fraction, strength and line profile, are consistent with expectations for LyC leakage from JADES-GS-z7-LA, its measured UV slope is not indicative of LyC leakage. In light of these findings, we conclude that the current LyC \fesc\ from JADES-GS-z7-LA is not necessarily substantially high. JADES-GS-z7-LA could have had significant LyC leakage in its past \citep[e.g. a `remnant' leaker from][]{kat23a}, but based on currently available data there is no reliable way to infer this. The current or past LyC leakage from JADES-GS-z7-LA has bearing on its ability to inflate an ionized bubble large enough to explain the observed \lya\ emission, and in the next section we investigate this in more detail.

\subsection{The nature of the ionized bubble surrounding JADES-GS-z7-LA}
To be able to observe bright \lya\ line emission (with high \fesc(\lya)) from a galaxy situated deep within the epoch of reionization, very close to the systemic redshift (e.g. $\Delta v < 150\,$\kms), the presence of a large ionized bubble surrounding the line-emitting source is required \citep[e.g.,][]{mir98}. The size of the ionized bubble that a high-redshift galaxy is capable of ionizing depends on both the galaxy properties as well as the global state of the IGM at that epoch \citep{hai97, fur06}. The main parameters responsible for controlling the size of this bubble are a galaxy's UV luminosity, its average ionizing photon production efficiency, \xiion, and the average fraction of LyC photons that manage to escape out of the galaxy, \fesc\ \citep[see][for recent examples]{end22a, cas22, tan23}. External factors that impact the size of the bubble are the average neutral hydrogen (H\textsc{i}) density and the Hubble parameter at the redshift in question, the H\textsc{i} clumping factor and the recombination coefficient.

The presence of a close companion source next to JADES-GS-z7-LA, which seemingly lies at the same redshift, may further help with the ionizing photon budget required to inflate a bubble of ionized gas large enough to let \lya\ emission through. Therefore, in this section we use simplified assumptions on the ionizing photon production histories of both JADES-GS-z7-LA and the companion to estimate whether one or both of these sources is/are capable of creating an ionized bubble large enough to explain the transmission of \lya\ emission.

To calculate the size of an ionized bubble that JADES-GS-z7-LA and/or its companion source could be capable of inflating, we mainly use the prescriptions laid out by \citet{mason20}, which are primarily geared towards the earlier, pre-overlap phase of reionization when ionized bubbles grew around individual sources. \citet{mason20} note that their calculations will effectively help establish lower limits on the size of the ionized bubbles around sources. To estimate the size of the ionized region needed to explain the observed \lya\ emission from the prescription above, a measurement of the observed velocity offset between \lya\ and the systemic redshift is needed, in addition to an estimate on the \lya\ transmission through the IGM at the source redshift. Although it is not possible to independently measure the \lya\ transmission for JADES-GS-z7-LA, the \lya\ escape fraction can provide valuable insights into the \lya\ transmission through the IGM under certain assumptions about the ISM.

The \lya\ transmission through the IGM will be equal to the \lya\ escape fraction if all of \lya\ emission produced along the line of sight also manages to escape from the ISM/CGM of the galaxy. In other words, any observed attenuation in the \lya\ flux is solely being caused by the IGM in this scenario. Owing to the faint (and consequently low-mass) nature of the LAE, it is not unreasonable to expect little to no gas in the surrounding CGM that could potentially attenuate \lya\ coming out of the galaxy. Further, the narrow line profile and the small velocity offset between the peak of \lya\ and the systemic redshift is indicative of low column densities of the neutral gas in the ISM of the LAE \citep[e.g.,][]{ver15}, which may result in minimal attenuation of \lya\ photons from the ISM. Therefore, it may not be unreasonable to assume that the \lya\ transmission through the IGM is the observed \lya\ escape fraction for JADES-GS-z7-LA.

Given the measured \fesc(\lya) of $0.45 - 1.00$ from a host of assumptions about the conditions in the H\,\textsc{ii} regions, Figure 1 of \citet{mason20} suggests that an observed \lya\ velocity offset of $\approx 113$\,\kms\ at $z\sim7$ will require an ionized bubble or sightline of size $\approx1.5-5$\,pMpc, or at least $>1.5$\,pMpc. 

Building on this, we now perform our calculations on the size of the ionized bubble that JADES-GS-z7-LA is capable of inflating. To do this, we fix the source redshift to $z_s = 7.278$ and we assume the IGM outside the ionized bubble to be neutral with a density equal to the expected density of neutral gas at $z_s = 7.278$ from the calculations of \citet{mad99}. In the simplest case, assuming a constant ionizing photon output from the sources, the size of the ionized bubble can be estimated using Equation (6) from \citet{mason20}. Other variables involved in this calculation are the escape fraction of ionizing photons and the time since the ionizing source has been switched on.

To estimate the ionizing photon output, we use Equation (9) from \citet{mason20}, which uses the absolute UV magnitude measured at $1500$\,\AA\ and the UV slope of the spectral energy distribution of the ionizing sources. Here we use the measured UV slope ($\beta$) for the value redward of the Lyman limit, and use a fiducial value of $\alpha = -2$ for the slope blueward of the Lyman limit.

\begin{figure}
    \centering
    \includegraphics[width=\linewidth]{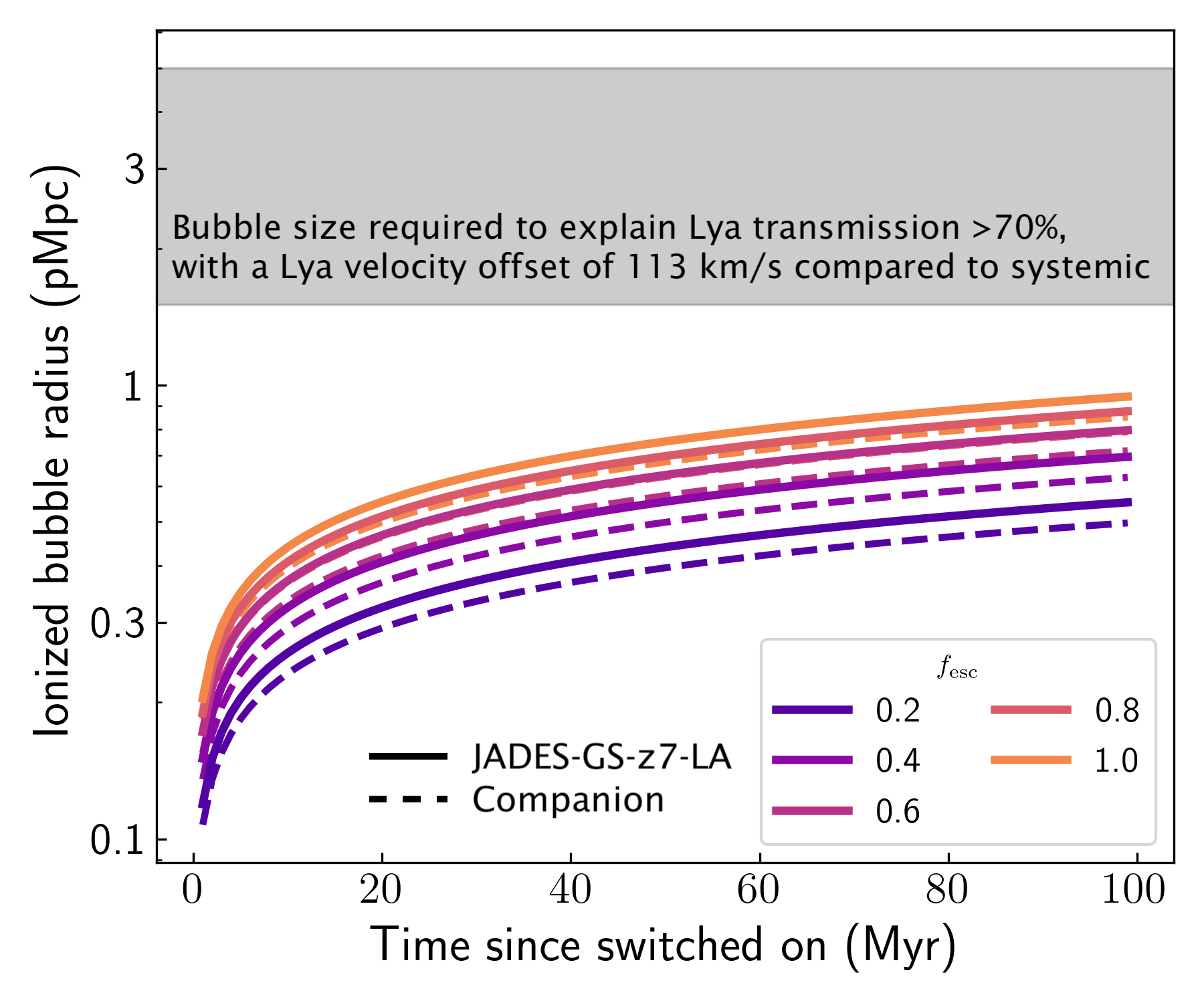}
    \caption{Predicted sized of an ionized bubble that JADES-GS-z7-LA (solid lines) and its companion source (dashed lines) as a function of time since star formation began for different LyC \fesc. We use very simplistic assumptions here, where the ionizing photon production rates are considered to be constant over the entire lifetime of the sources, set to the currently inferred value from the UV magnitude. We find that neither JADES-GS-z7-LA nor the companion are capable of inflating an ionized bubble of size $>1.5$\,pMpc (shaded region), even if they had \fesc(LyC) $=1.0$ over the past 100 Myr of their lifetime. The visibility of \lya\ from JADES-GS-z7-LA, therefore, suggests that these galaxies may be part of an ionized bubble inflated by other galaxies in the field, or be situated in an opportunistic sightline cleared of neutral hydrogen by intervening sources.}
    \label{fig:bubble}
\end{figure}

In Figure \ref{fig:bubble} we show the expected size of the bubble that JADES-GS-z7-LA (solid lines) and its companion (dashed lines) would be capable of inflating for a range of LyC escape fractions over a total switched on time of 30\,Myr, which is close to the maximum age of star formation activity that occurred in both galaxies based on best-fitting SEDs.
We find that neither JADES-GS-z7-LA nor its companion source seem to be able to inflate a bubble of radius $\gtrsim3$\,pMpc on their own, required for \lya\ emission to redshift out of resonance within $\sim110$\,km/s of the systemic redshift at $z\sim7$, fully accounting for the observed \lya\ emission seen from JADES-GS-z7-LA. Even with a constant \fesc $=1.0$ over $\sim100$\,Myr, the estimated bubble size would fall substantially short of the $>1.5$\,pMpc requirement. This is perhaps not surprising, as the sources reported in this study are several orders of magnitude fainter than those reported by \citet{end22a} and \citet{cas22} for example, that are capable of ionizing larger bubbles themselves.

Qualitatively speaking, given the relative faintness of both sources, both of them working in combination to ionize their immediate surroundings would also not be enough to inflate a bubble of radius $>1.5$\,pMpc. Such a calculation would require careful radiative transfer modeling of the impact of ionizing photons produced by two nearby sources in a neutral medium, which is beyond the scope of this work.

Therefore, given the extreme continuum faintness of JADES-GS-z7-LA and the high EW of \lya\ emission, it is unlikely that either it or the companion source would be able to produce an ionized bubble that is large enough to explain the observed \lya\ emission. We note here that assuming that a single ionized bubble is responsible for the transmission of the observed \lya\ emission may lead to overestimates on the size required. Instead, an ionized line of sight towards an LAE at $z>6$ can also achieve the required ionization of the IGM along the given line of sight, without requiring a single large ionized bubble. 

A possible explanation to explain the decreased neutral fraction around JADES-GS-z7-LA and its companion source could be that these galaxies may be part of a larger overdensity of galaxies at $z\approx7.3$ in the GOODS-S field, as a result of which the IGM neutral fraction may be lower than average \citep[e.g.,][]{tra23}. This would indicate that the ionized bubbles from sources in the field at $z\approx7.3$ have already begun to overlap \citep[e.g.,][]{wit23b}.

\section{Conclusions}
\label{sec:conclusions}

In this paper, we have presented the discovery of extremely high equivalent width \lya\ emission (\lya\ EW $=388$\,\AA) from a continuum-faint galaxy ($M_{\rm{UV}} \approx -17.0)$ at $z=7.278$, called JADES-GS-z7-LA, using \emph{JWST}/NIRSpec. The \lya\ emission is clearly detected both in low resolution ($R\sim100$) PRISM as well as medium resolution ($R\sim1000$) G140M grating spectra. Rest-frame optical emission lines such as \oii, \hb\ and \oiii\ are also well-detected in the PRISM spectrum, leading to an accurate spectroscopic redshift.

The \lya\ emission is remarkably bright and narrow (LSF-deconvolved FWHM $= 383.9 \pm 56.2$\,\kms). The velocity offset of the peak of the \lya\ line compared to the systemic velocity is only $113.3 \pm 80.0$\,\kms, suggesting the existence of low column density channels in the neutral gas as well as an ionized bubble around the galaxy. The high EW \lya\ emission suggests a fairly dust-free environment along the line of sight, which is consistent with the limits on the Balmer decrements placed using \hb\ and the non-detection of \hg.

Deep NIRCam imaging has additionally revealed the presence of a nearby ($\sim0.23''$ away) companion source, which was not covered in the MSA shutter placement that led to the spectrum of JADES-GS-z7-LA. The photometric properties of the companion source, however, are consistent with that of JADES-GS-z7-LA, with a strong excess seen in the F410M NIRCam medium band tracing \oiii+\hb\ at $z\approx7.3$. This excess is understandably also seen in JADES-GS-z7-LA. There is also an excess in the F115W broad band from the companion source, consistent with \lya\ emission at the same redshift. 

Considering the evidence that the companion also lies at $z\approx7.3$, we find that the best-fitting SED is able to reproduce the photometry of the companion reasonably well when the redshift is fixed to that of JADES-GS-z7-LA. The best-fitting SED parameters indicate that JADES-GS-z7-LA has had a ``bursty'' star formation history, with $\approx53\%$ of its stellar mass formed in a recent burst $\approx2.2$\,Myr long. The companion source on the other hand has only assembled $\approx14\%$ of its stellar mass over a recent burst lasting $\approx3$\,Myr, with the bulk of its stellar mass formed over a longer period.

We report that under assumptions of no dust and Case B recombination, the inferred escape fraction of \lya\ photons from JADES-GS-z7-LA is in the range $0.7-1.0$. This suggests that in addition to a lack of substantial neutral gas and dust within JADES-GS-z7-LA, it must also exist in a fairly ionized bubble in an otherwise neutral intergalactic medium. We investigate signs of significant Lyman continuum (LyC) photon escape from JADES-GS-z7-LA based on its spectroscopic and photometric properties and find that although certain spectroscopic indicators are consistent with known LyC leakers at lower redshifts, its relatively shallow UV slope ($\beta \approx-2.0$) is incompatible with models suggestive of significant LyC leakage. A shallow UV slope may trace significant nebular continuum emission, resulting in a large fraction of LyC photons being absorbed by the gas instead of escaping out of the galaxy.

Using simple assumptions on the history of ionizing photon production and escape, we find that JADES-GS-z7-LA and its companion source would not be able to individually inflate an ionized bubble large enough ($>1.5$\,pMpc) that is needed to explain the observed \lya\ emission strength and velocity offset, even if they have had a ionizing photon escape fraction, \fesc(LyC) $=1.0$ continuously over a period of $100$\,Myr. Based on this we argue that both JADES-GS-z7-LA and its companion, if confirmed to lie at the same redshift, may be part of a larger overdensity in the field at $z\approx7.3$, tracing an epoch when the ionized bubbles have already begun to overlap in the field.

These observations demonstrate the incredibly transformative power of \emph{JWST} NIRSpec and NIRCam observations in discovering strong emission lines from distant galaxies, leading to an increased understanding of the relatively poorly understood contribution of fainter star-forming galaxies towards cosmic reionization.

\begin{acknowledgements}
We thank the referee for constructive feedback and comments that helped improve the quality of this work. AS thanks Harley Katz, Richard Ellis, Anne Jaskot and Charlotte Mason for insightful discussions. AS, AJB, AJC, GCJ and JC acknowledge funding from the ``FirstGalaxies'' Advanced Grant from the European Research Council (ERC) under the European Union’s Horizon 2020 research and innovation programme (Grant agreement No. 789056). BER, BDJ, EE, and MR acknowledge support from the NIRCam Science Team contract to the University of Arizona, NAS5-02015. FDE, MC, TJL, JW, LS and JS acknowledge support by the Science and Technology Facilities Council (STFC), ERC Advanced Grant 695671 "QUENCH". JW also acknowledges support from the Fondation MERAC. SC acknowledges support by European Union's HE ERC Starting Grant No. 101040227 - WINGS. ECL acknowledges support of an STFC Webb Fellowship (ST/W001438/1). SA and BRDP acknowledge support from the research project PID2021-127718NB- I00 of the Spanish Ministry of Science and Innovation/State Agency of Research (MICIN/AEI). RS acknowledges support from a STFC Ernest Rutherford Fellowship (ST/S004831/1). H\"{U} gratefully acknowledges support by the Isaac Newton Trust and by the Kavli Foundation through a Newton-Kavli Junior Fellowship. DJE is supported as a Simons Investigator and by JWST/NIRCam contract to the University of Arizona, NAS5- 02015. The work of CCW is supported by NOIR-Lab, which is managed by the Association of Universities for Research in Astronomy (AURA) under a cooperative agreement with the National Science Foundation. RB acknowledges sup- port from an STFC Ernest Rutherford Fellowship (ST/T003596/1). KB acknowledges support from the Australian Research Council center of Excellence for All Sky Astrophysics in 3 Dimensions (ASTRO 3D), through project number CE170100013.

This work is based on observations made with the NASA/ESA/CSA James Webb Space Telescope. The data were obtained from the Mikulski Archive for Space Telescopes at the Space Telescope Science Institute, which is operated by the Association of Universities for Research in Astronomy, Inc., under NASA contract NAS 5-03127 for JWST. These observations are associated with program \#1180 and \#1210.

\end{acknowledgements}

\bibliographystyle{aa}
\bibliography{jades_lae_z7_bib}

\begin{thebibliography}{123}
\expandafter\ifx\csname natexlab\endcsname\relax\def\natexlab#1{#1}\fi

\bibitem[{{Arellano-C{\'o}rdova} {et~al.}(2022){Arellano-C{\'o}rdova}, {Berg},
  {Chisholm}, {Arrabal Haro}, {Dickinson}, {Finkelstein}, {Leclercq}, {Rogers},
  {Simons}, {Skillman}, {Trump}, \& {Kartaltepe}}]{are22}
{Arellano-C{\'o}rdova}, K.~Z., {Berg}, D.~A., {Chisholm}, J., {et~al.} 2022,
  \apjl, 940, L23

\bibitem[{{Atek} {et~al.}(2008){Atek}, {Kunth}, {Hayes}, {{\"O}stlin}, \&
  {Mas-Hesse}}]{ate08}
{Atek}, H., {Kunth}, D., {Hayes}, M., {{\"O}stlin}, G., \& {Mas-Hesse}, J.~M.
  2008, \aap, 488, 491

\bibitem[{{Birkmann} {et~al.}(2022){Birkmann}, {Giardino}, {Sirianni},
  {Ferruit}, {Rauscher}, {Alves de Oliveira}, {B{\"o}ker}, {Kumari},
  {L{\"u}tzgendorf}, {Manjavacas}, {Proffitt}, {Rawle}, {te Plate}, \&
  {Zeidler}}]{bir22}
{Birkmann}, S.~M., {Giardino}, G., {Sirianni}, M., {et~al.} 2022, in Society of
  Photo-Optical Instrumentation Engineers (SPIE) Conference Series, Vol. 12180,
  Space Telescopes and Instrumentation 2022: Optical, Infrared, and Millimeter
  Wave, ed. L.~E. {Coyle}, S.~{Matsuura}, \& M.~D. {Perrin}, 121802P

\bibitem[{{B{\"o}ker} {et~al.}(2023){B{\"o}ker}, {Beck}, {Birkmann},
  {Giardino}, {Keyes}, {Kumari}, {Muzerolle}, {Rawle}, {Zeidler}, {Abul-Huda},
  {Alves de Oliveira}, {Arribas}, {Bechtold}, {Bhatawdekar}, {Bonaventura},
  {Bunker}, {Cameron}, {Carniani}, {Charlot}, {Curti}, {Espinoza}, {Ferruit},
  {Franx}, {Jakobsen}, {Karakla}, {L{\'o}pez-Caniego}, {L{\"u}tzgendorf},
  {Maiolino}, {Manjavacas}, {Marston}, {Moseley}, {Ogle}, {Perna},
  {Pe{\~n}a-Guerrero}, {Pirzkal}, {Plesha}, {Proffitt}, {Rauscher}, {Rix},
  {Rodr{\'\i}guez del Pino}, {Rustamkulov}, {Sabbi}, {Sing}, {Sirianni}, {te
  Plate}, {{\'U}beda}, {Wahlgren}, {Wislowski}, {Wu}, \& {Willott}}]{bok23}
{B{\"o}ker}, T., {Beck}, T.~L., {Birkmann}, S.~M., {et~al.} 2023, \pasp, 135,
  038001

\bibitem[{{Bolan} {et~al.}(2022){Bolan}, {Lemaux}, {Mason}, {Brada{\v{c}}},
  {Treu}, {Strait}, {Pelliccia}, {Pentericci}, \& {Malkan}}]{bol22}
{Bolan}, P., {Lemaux}, B.~C., {Mason}, C., {et~al.} 2022, \mnras, 517, 3263

\bibitem[{{Bouwens} {et~al.}(2022){Bouwens}, {Illingworth}, {Ellis}, {Oesch},
  \& {Stefanon}}]{bou22a}
{Bouwens}, R.~J., {Illingworth}, G., {Ellis}, R.~S., {Oesch}, P., \&
  {Stefanon}, M. 2022, \apj, 940, 55

\bibitem[{{Bruzual} \& {Charlot}(2003)}]{bc03}
{Bruzual}, G. \& {Charlot}, S. 2003, \mnras, 344, 1000

\bibitem[{{Bunker} {et~al.}(2023{\natexlab{a}}){Bunker}, {Cameron},
  {Curtis-Lake}, {Jakobsen}, {Carniani}, {Curti}, {Witstok}, {Maiolino},
  {D'Eugenio}, {Looser}, {Willott}, {Bonaventura}, {Hainline}, {Uebler},
  {Willmer}, {Saxena}, {Smit}, {Alberts}, {Arribas}, {Baker}, {Baum},
  {Bhatawdekar}, {Bowler}, {Boyett}, {Charlot}, {Chen}, {Chevallard},
  {Circosta}, {DeCoursey}, {de Graaff}, {Egami}, {Eisenstein}, {Endsley},
  {Ferruit}, {Giardino}, {Hausen}, {Helton}, {Hviding}, {Ji}, {Johnson},
  {Jones}, {Kumari}, {Laseter}, {Luetzgendorf}, {Maseda}, {Nelson}, {Parlanti},
  {Perna}, {Rawle}, {Rix}, {Rieke}, {Robertson}, {Rodriguez Del Pino},
  {Sandles}, {Scholtz}, {Sharpe}, {Skarbinski}, {Stark}, {Sun}, {Tacchella},
  {Topping}, {Villanueva}, {Wallace}, {Williams}, \& {Woodrum}}]{bun23b}
{Bunker}, A.~J., {Cameron}, A.~J., {Curtis-Lake}, E., {et~al.}
  2023{\natexlab{a}}, arXiv e-prints, arXiv:2306.02467

\bibitem[{{Bunker} {et~al.}(2023{\natexlab{b}}){Bunker}, {Saxena}, {Cameron},
  {Willott}, {Curtis-Lake}, {Jakobsen}, {Carniani}, {Smit}, {Maiolino},
  {Witstok}, {Curti}, {D'Eugenio}, {Jones}, {Ferruit}, {Arribas}, {Charlot},
  {Chevallard}, {Giardino}, {de Graaff}, {Looser}, {Luetzgendorf}, {Maseda},
  {Rawle}, {Rix}, {Rodriguez Del Pino}, {Alberts}, {Egami}, {Eisenstein},
  {Endsley}, {Hainline}, {Hausen}, {Johnson}, {Rieke}, {Rieke}, {Robertson},
  {Shivaei}, {Stark}, {Sun}, {Tacchella}, {Tang}, {Williams}, {Willmer},
  {Baker}, {Baum}, {Bhatawdekar}, {Bowler}, {Boyett}, {Chen}, {Circosta},
  {Helton}, {Ji}, {Lyu}, {Nelson}, {Parlanti}, {Perna}, {Sandles}, {Scholtz},
  {Suess}, {Topping}, {Uebler}, {Wallace}, \& {Whitler}}]{bun23}
{Bunker}, A.~J., {Saxena}, A., {Cameron}, A.~J., {et~al.} 2023{\natexlab{b}},
  arXiv e-prints, arXiv:2302.07256

\bibitem[{{Cameron} {et~al.}(2023){Cameron}, {Saxena}, {Bunker}, {D'Eugenio},
  {Carniani}, {Maiolino}, {Curtis-Lake}, {Ferruit}, {Jakobsen}, {Arribas},
  {Bonaventura}, {Charlot}, {Chevallard}, {Curti}, {Looser}, {Maseda}, {Rawle},
  {Rodr{\'\i}guez Del Pino}, {Smit}, {{\"U}bler}, {Willott}, {Witstok},
  {Egami}, {Eisenstein}, {Johnson}, {Hainline}, {Rieke}, {Robertson}, {Stark},
  {Tacchella}, {Williams}, {Bhatawdekar}, {Bowler}, {Boyett}, {Circosta},
  {Helton}, {Jones}, {Kumari}, {Ji}, {Nelson}, {Parlanti}, {Sandles},
  {Scholtz}, \& {Sun}}]{cam23}
{Cameron}, A.~J., {Saxena}, A., {Bunker}, A.~J., {et~al.} 2023, arXiv e-prints,
  arXiv:2302.04298

\bibitem[{{Carnall} {et~al.}(2018){Carnall}, {McLure}, {Dunlop}, \&
  {Dav{\'e}}}]{car18}
{Carnall}, A.~C., {McLure}, R.~J., {Dunlop}, J.~S., \& {Dav{\'e}}, R. 2018,
  \mnras, 480, 4379

\bibitem[{{Carniani} {et~al.}(2017){Carniani}, {Maiolino}, {Pallottini},
  {Vallini}, {Pentericci}, {Ferrara}, {Castellano}, {Vanzella}, {Grazian},
  {Gallerani}, {Santini}, {Wagg}, \& {Fontana}}]{car17}
{Carniani}, S., {Maiolino}, R., {Pallottini}, A., {et~al.} 2017, \aap, 605, A42

\bibitem[{{Castellano} {et~al.}(2022){Castellano}, {Pentericci}, {Cupani},
  {Curtis-Lake}, {Vanzella}, {Amor{\'\i}n}, {Belfiori}, {Calabr{\`o}},
  {Carniani}, {Charlot}, {Chevallard}, {Dayal}, {Dickinson}, {Ferrara},
  {Fontana}, {Giallongo}, {Hutter}, {Merlin}, {Paris}, \& {Santini}}]{cas22}
{Castellano}, M., {Pentericci}, L., {Cupani}, G., {et~al.} 2022, \aap, 662,
  A115

\bibitem[{{Castellano} {et~al.}(2018){Castellano}, {Pentericci}, {Vanzella},
  {Marchi}, {Fontana}, {Dayal}, {Ferrara}, {Hutter}, {Carniani}, {Cristiani},
  {Dickinson}, {Gallerani}, {Giallongo}, {Giavalisco}, {Grazian}, {Maiolino},
  {Merlin}, {Paris}, {Pilo}, \& {Santini}}]{cas18}
{Castellano}, M., {Pentericci}, L., {Vanzella}, E., {et~al.} 2018, \apjl, 863,
  L3

\bibitem[{{Charlot} \& {Longhetti}(2001)}]{cha01}
{Charlot}, S. \& {Longhetti}, M. 2001, \mnras, 323, 887

\bibitem[{{Chisholm} {et~al.}(2022){Chisholm}, {Saldana-Lopez}, {Flury},
  {Schaerer}, {Jaskot}, {Amor{\'\i}n}, {Atek}, {Finkelstein}, {Fleming},
  {Ferguson}, {Fern{\'a}ndez}, {Giavalisco}, {Hayes}, {Heckman}, {Henry}, {Ji},
  {Marques-Chaves}, {Mauerhofer}, {McCandliss}, {Oey}, {{\"O}stlin},
  {Rutkowski}, {Scarlata}, {Thuan}, {Trebitsch}, {Wang}, {Worseck}, \&
  {Xu}}]{chi22}
{Chisholm}, J., {Saldana-Lopez}, A., {Flury}, S., {et~al.} 2022, \mnras, 517,
  5104

\bibitem[{{Choustikov} {et~al.}(2023){Choustikov}, {Katz}, {Saxena}, {Cameron},
  {Devriendt}, {Slyz}, {Rosdahl}, {Blaizot}, \& {Michel-Dansac}}]{cho23}
{Choustikov}, N., {Katz}, H., {Saxena}, A., {et~al.} 2023, arXiv e-prints,
  arXiv:2304.08526

\bibitem[{{Cullen} {et~al.}(2023){Cullen}, {McLure}, {McLeod}, {Dunlop},
  {Donnan}, {Carnall}, {Bowler}, {Begley}, {Hamadouche}, \& {Stanton}}]{cul23}
{Cullen}, F., {McLure}, R.~J., {McLeod}, D.~J., {et~al.} 2023, \mnras, 520, 14

\bibitem[{{Curti} {et~al.}(2023){Curti}, {D'Eugenio}, {Carniani}, {Maiolino},
  {Sandles}, {Witstok}, {Baker}, {Bennett}, {Piotrowska}, {Tacchella},
  {Charlot}, {Nakajima}, {Maheson}, {Mannucci}, {Amiri}, {Arribas}, {Belfiore},
  {Bonaventura}, {Bunker}, {Chevallard}, {Cresci}, {Curtis-Lake},
  {Hayden-Pawson}, {Jones}, {Kumari}, {Laseter}, {Looser}, {Marconi}, {Maseda},
  {Scholtz}, {Smit}, {{\"U}bler}, \& {Wallace}}]{cur23}
{Curti}, M., {D'Eugenio}, F., {Carniani}, S., {et~al.} 2023, \mnras, 518, 425

\bibitem[{{Curtis-Lake} {et~al.}(2023){Curtis-Lake}, {Carniani}, {Cameron},
  {Charlot}, {Jakobsen}, {Maiolino}, {Bunker}, {Witstok}, {Smit}, {Chevallard},
  {Willott}, {Ferruit}, {Arribas}, {Bonaventura}, {Curti}, {D'Eugenio},
  {Franx}, {Giardino}, {Looser}, {L{\"u}tzgendorf}, {Maseda}, {Rawle}, {Rix},
  {Rodr{\'\i}guez del Pino}, {{\"U}bler}, {Sirianni}, {Dressler}, {Egami},
  {Eisenstein}, {Endsley}, {Hainline}, {Hausen}, {Johnson}, {Rieke},
  {Robertson}, {Shivaei}, {Stark}, {Tacchella}, {Williams}, {Willmer},
  {Bhatawdekar}, {Bowler}, {Boyett}, {Chen}, {de Graaff}, {Helton}, {Hviding},
  {Jones}, {Kumari}, {Lyu}, {Nelson}, {Perna}, {Sandles}, {Saxena}, {Suess},
  {Sun}, {Topping}, {Wallace}, \& {Whitler}}]{curtis23}
{Curtis-Lake}, E., {Carniani}, S., {Cameron}, A., {et~al.} 2023, Nature
  Astronomy, 7, 622

\bibitem[{{Dayal} \& {Ferrara}(2018)}]{day18}
{Dayal}, P. \& {Ferrara}, A. 2018, \physrep, 780, 1

\bibitem[{{Dijkstra} {et~al.}(2016){Dijkstra}, {Gronke}, \&
  {Venkatesan}}]{dij16}
{Dijkstra}, M., {Gronke}, M., \& {Venkatesan}, A. 2016, \apj, 828, 71

\bibitem[{{Donnan} {et~al.}(2023){Donnan}, {McLeod}, {Dunlop}, {McLure},
  {Carnall}, {Begley}, {Cullen}, {Hamadouche}, {Bowler}, {Magee}, {McCracken},
  {Milvang-Jensen}, {Moneti}, \& {Targett}}]{don23}
{Donnan}, C.~T., {McLeod}, D.~J., {Dunlop}, J.~S., {et~al.} 2023, \mnras, 518,
  6011

\bibitem[{{Dopita} \& {Sutherland}(2003)}]{dop03}
{Dopita}, M.~A. \& {Sutherland}, R.~S. 2003, {Astrophysics of the diffuse
  universe}

\bibitem[{{Eisenstein} {et~al.}(2023){Eisenstein}, {Willott}, {Alberts},
  {Arribas}, {Bonaventura}, {Bunker}, {Cameron}, {Carniani}, {Charlot},
  {Curtis-Lake}, {D'Eugenio}, {Endsley}, {Ferruit}, {Giardino}, {Hainline},
  {Hausen}, {Jakobsen}, {Johnson}, {Maiolino}, {Rieke}, {Rieke}, {Rix},
  {Robertson}, {Stark}, {Tacchella}, {Williams}, {Willmer}, {Baker}, {Baum},
  {Bhatawdekar}, {Boyett}, {Chen}, {Chevallard}, {Circosta}, {Curti},
  {Danhaive}, {DeCoursey}, {de Graaff}, {Dressler}, {Egami}, {Helton},
  {Hviding}, {Ji}, {Jones}, {Kumari}, {L{\"u}tzgendorf}, {Laseter}, {Looser},
  {Lyu}, {Maseda}, {Nelson}, {Parlanti}, {Perna}, {Pusk{\'a}s}, {Rawle},
  {Rodr{\'\i}guez Del Pino}, {Sandles}, {Saxena}, {Scholtz}, {Sharpe},
  {Shivaei}, {Silcock}, {Simmonds}, {Skarbinski}, {Smit}, {Stone}, {Suess},
  {Sun}, {Tang}, {Topping}, {{\"U}bler}, {Villanueva}, {Wallace}, {Whitler},
  {Witstok}, \& {Woodrum}}]{eis23}
{Eisenstein}, D.~J., {Willott}, C., {Alberts}, S., {et~al.} 2023, arXiv
  e-prints, arXiv:2306.02465

\bibitem[{{Endsley} \& {Stark}(2022)}]{end22a}
{Endsley}, R. \& {Stark}, D.~P. 2022, \mnras, 511, 6042

\bibitem[{{Endsley} {et~al.}(2022){Endsley}, {Stark}, {Bouwens}, {Schouws},
  {Smit}, {Stefanon}, {Inami}, {Bowler}, {Oesch}, {Gonzalez}, {Aravena}, {da
  Cunha}, {Dayal}, {Ferrara}, {Graziani}, {Nanayakkara}, {Pallottini},
  {Schneider}, {Sommovigo}, {Topping}, {van der Werf}, \& {Hutter}}]{end22c}
{Endsley}, R., {Stark}, D.~P., {Bouwens}, R.~J., {et~al.} 2022, \mnras, 517,
  5642

\bibitem[{{Endsley} {et~al.}(2021){Endsley}, {Stark}, {Chevallard}, \&
  {Charlot}}]{end21}
{Endsley}, R., {Stark}, D.~P., {Chevallard}, J., \& {Charlot}, S. 2021, \mnras,
  500, 5229

\bibitem[{{Endsley} {et~al.}(2023){Endsley}, {Stark}, {Whitler}, {Topping},
  {Chen}, {Plat}, {Chisholm}, \& {Charlot}}]{end23}
{Endsley}, R., {Stark}, D.~P., {Whitler}, L., {et~al.} 2023, \mnras, 524, 2312

\bibitem[{{Faisst}(2016)}]{fai16}
{Faisst}, A.~L. 2016, \apj, 829, 99

\bibitem[{{Ferland} {et~al.}(2013){Ferland}, {Porter}, {van Hoof}, {Williams},
  {Abel}, {Lykins}, {Shaw}, {Henney}, \& {Stancil}}]{fer13}
{Ferland}, G.~J., {Porter}, R.~L., {van Hoof}, P.~A.~M., {et~al.} 2013, \rmxaa,
  49, 137

\bibitem[{{Ferruit} {et~al.}(2022){Ferruit}, {Jakobsen}, {Giardino}, {Rawle},
  {Alves de Oliveira}, {Arribas}, {Beck}, {Birkmann}, {B{\"o}ker}, {Bunker},
  {Charlot}, {de Marchi}, {Franx}, {Henry}, {Karakla}, {Kassin}, {Kumari},
  {L{\'o}pez-Caniego}, {L{\"u}tzgendorf}, {Maiolino}, {Manjavacas}, {Marston},
  {Moseley}, {Muzerolle}, {Pirzkal}, {Rauscher}, {Rix}, {Sabbi}, {Sirianni},
  {te Plate}, {Valenti}, {Willott}, \& {Zeidler}}]{fer22}
{Ferruit}, P., {Jakobsen}, P., {Giardino}, G., {et~al.} 2022, \aap, 661, A81

\bibitem[{{Finkelstein} {et~al.}(2011){Finkelstein}, {Cohen}, {Moustakas},
  {Malhotra}, {Rhoads}, \& {Papovich}}]{fin10}
{Finkelstein}, S.~L., {Cohen}, S.~H., {Moustakas}, J., {et~al.} 2011, \apj,
  733, 117

\bibitem[{{Finkelstein} {et~al.}(2019){Finkelstein}, {D'Aloisio},
  {Paardekooper}, {Ryan}, {Behroozi}, {Finlator}, {Livermore}, {Upton
  Sanderbeck}, {Dalla Vecchia}, \& {Khochfar}}]{fin19}
{Finkelstein}, S.~L., {D'Aloisio}, A., {Paardekooper}, J.-P., {et~al.} 2019,
  \apj, 879, 36

\bibitem[{{Fujimoto} {et~al.}(2023){Fujimoto}, {Arrabal Haro}, {Dickinson},
  {Finkelstein}, {Kartaltepe}, {Larson}, {Burgarella}, {Bagley}, {Behroozi},
  {Chworowsky}, {Hirschmann}, {Trump}, {Wilkins}, {Yung}, {Koekemoer},
  {Papovich}, {Pirzkal}, {Ferguson}, {Fontana}, {Grogin}, {Grazian}, {Kewley},
  {Kocevski}, {Lotz}, {Pentericci}, {Ravindranath}, {Somerville}, {Wilkins},
  {Amor{\'\i}n}, {Backhaus}, {Calabr{\`o}}, {Casey}, {Cooper}, {Fern{\'a}ndez},
  {Franco}, {Giavalisco}, {Hathi}, {Harish}, {Hutchison}, {Iyer}, {Jung},
  {Lucas}, \& {Zavala}}]{fuj23}
{Fujimoto}, S., {Arrabal Haro}, P., {Dickinson}, M., {et~al.} 2023, \apjl, 949,
  L25

\bibitem[{{Fuller} {et~al.}(2020){Fuller}, {Lemaux}, {Brada{\v{c}}}, {Hoag},
  {Schmidt}, {Huang}, {Strait}, {Mason}, {Treu}, {Pentericci}, {Trenti},
  {Henry}, \& {Malkan}}]{ful20}
{Fuller}, S., {Lemaux}, B.~C., {Brada{\v{c}}}, M., {et~al.} 2020, \apj, 896,
  156

\bibitem[{{Furlanetto} {et~al.}(2006){Furlanetto}, {Zaldarriaga}, \&
  {Hernquist}}]{fur06}
{Furlanetto}, S.~R., {Zaldarriaga}, M., \& {Hernquist}, L. 2006, \mnras, 365,
  1012

\bibitem[{{Furusawa} {et~al.}(2016){Furusawa}, {Kashikawa}, {Kobayashi},
  {Dunlop}, {Shimasaku}, {Takata}, {Sekiguchi}, {Naito}, {Furusawa}, {Ouchi},
  {Nakata}, {Yasuda}, {Okura}, {Taniguchi}, {Yamada}, {Kajisawa}, {Fynbo}, \&
  {Le F{\`e}vre}}]{fur16}
{Furusawa}, H., {Kashikawa}, N., {Kobayashi}, M. A.~R., {et~al.} 2016, \apj,
  822, 46

\bibitem[{{Gazagnes} {et~al.}(2020){Gazagnes}, {Chisholm}, {Schaerer},
  {Verhamme}, \& {Izotov}}]{gaz20}
{Gazagnes}, S., {Chisholm}, J., {Schaerer}, D., {Verhamme}, A., \& {Izotov}, Y.
  2020, \aap, 639, A85

\bibitem[{{Gordon} {et~al.}(2003){Gordon}, {Clayton}, {Misselt}, {Landolt}, \&
  {Wolff}}]{gor03}
{Gordon}, K.~D., {Clayton}, G.~C., {Misselt}, K.~A., {Landolt}, A.~U., \&
  {Wolff}, M.~J. 2003, \apj, 594, 279

\bibitem[{{Haiman} \& {Loeb}(1997)}]{hai97}
{Haiman}, Z. \& {Loeb}, A. 1997, \apj, 483, 21

\bibitem[{{Hashimoto} {et~al.}(2019){Hashimoto}, {Inoue}, {Mawatari}, {Tamura},
  {Matsuo}, {Furusawa}, {Harikane}, {Shibuya}, {Knudsen}, {Kohno}, {Ono},
  {Zackrisson}, {Okamoto}, {Kashikawa}, {Oesch}, {Ouchi}, {Ota}, {Shimizu},
  {Taniguchi}, {Umehata}, \& {Watson}}]{has19}
{Hashimoto}, T., {Inoue}, A.~K., {Mawatari}, K., {et~al.} 2019, \pasj, 71, 71

\bibitem[{{Heckman} {et~al.}(2001){Heckman}, {Sembach}, {Meurer}, {Leitherer},
  {Calzetti}, \& {Martin}}]{hec01}
{Heckman}, T.~M., {Sembach}, K.~R., {Meurer}, G.~R., {et~al.} 2001, \apj, 558,
  56

\bibitem[{{Hoag} {et~al.}(2019){Hoag}, {Brada{\v{c}}}, {Huang}, {Mason},
  {Treu}, {Schmidt}, {Trenti}, {Strait}, {Lemaux}, {Finney}, \&
  {Paddock}}]{hoa19}
{Hoag}, A., {Brada{\v{c}}}, M., {Huang}, K., {et~al.} 2019, \apj, 878, 12

\bibitem[{{Hu} {et~al.}(2021){Hu}, {Wang}, {Infante}, {Rhoads}, {Zheng},
  {Yang}, {Malhotra}, {Barrientos}, {Jiang}, {Gonz{\'a}lez-L{\'o}pez},
  {Prieto}, {Perez}, {Hibon}, {Galaz}, {Coughlin}, {Harish}, {Kong}, {Kang},
  {Khostovan}, {Pharo}, {Valdes}, {Wold}, {Walker}, \& {Zheng}}]{hu21}
{Hu}, W., {Wang}, J., {Infante}, L., {et~al.} 2021, Nature Astronomy, 5, 485

\bibitem[{{Inoue} {et~al.}(2014){Inoue}, {Shimizu}, {Iwata}, \&
  {Tanaka}}]{ino14}
{Inoue}, A.~K., {Shimizu}, I., {Iwata}, I., \& {Tanaka}, M. 2014, \mnras, 442,
  1805

\bibitem[{{Izotov} {et~al.}(2018){Izotov}, {Schaerer}, {Worseck}, {Guseva},
  {Thuan}, {Verhamme}, {Orlitov{\'a}}, \& {Fricke}}]{izo18a}
{Izotov}, Y.~I., {Schaerer}, D., {Worseck}, G., {et~al.} 2018, \mnras, 474,
  4514

\bibitem[{{Izotov} {et~al.}(2021){Izotov}, {Worseck}, {Schaerer}, {Guseva},
  {Chisholm}, {Thuan}, {Fricke}, \& {Verhamme}}]{izo21}
{Izotov}, Y.~I., {Worseck}, G., {Schaerer}, D., {et~al.} 2021, \mnras, 503,
  1734

\bibitem[{{Jakobsen} {et~al.}(2022){Jakobsen}, {Ferruit}, {Alves de Oliveira},
  {Arribas}, {Bagnasco}, {Barho}, {Beck}, {Birkmann}, {B{\"o}ker}, {Bunker},
  {Charlot}, {de Jong}, {de Marchi}, {Ehrenwinkler}, {Falcolini}, {Fels},
  {Franx}, {Franz}, {Funke}, {Giardino}, {Gnata}, {Holota}, {Honnen}, {Jensen},
  {Jentsch}, {Johnson}, {Jollet}, {Karl}, {Kling}, {K{\"o}hler}, {Kolm},
  {Kumari}, {Lander}, {Lemke}, {L{\'o}pez-Caniego}, {L{\"u}tzgendorf},
  {Maiolino}, {Manjavacas}, {Marston}, {Maschmann}, {Maurer}, {Messerschmidt},
  {Moseley}, {Mosner}, {Mott}, {Muzerolle}, {Pirzkal}, {Pittet}, {Plitzke},
  {Posselt}, {Rapp}, {Rauscher}, {Rawle}, {Rix}, {R{\"o}del}, {Rumler},
  {Sabbi}, {Salvignol}, {Schmid}, {Sirianni}, {Smith}, {Strada}, {te Plate},
  {Valenti}, {Wettemann}, {Wiehe}, {Wiesmayer}, {Willott}, {Wright}, {Zeidler},
  \& {Zincke}}]{jak22}
{Jakobsen}, P., {Ferruit}, P., {Alves de Oliveira}, C., {et~al.} 2022, \aap,
  661, A80

\bibitem[{{Jung} {et~al.}(2020){Jung}, {Finkelstein}, {Dickinson}, {Hutchison},
  {Larson}, {Papovich}, {Pentericci}, {Straughn}, {Guo}, {Malhotra}, {Rhoads},
  {Song}, {Tilvi}, \& {Wold}}]{jun20}
{Jung}, I., {Finkelstein}, S.~L., {Dickinson}, M., {et~al.} 2020, \apj, 904,
  144

\bibitem[{{Katz} {et~al.}(2023{\natexlab{a}}){Katz}, {Saxena}, {Cameron},
  {Carniani}, {Bunker}, {Arribas}, {Bhatawdekar}, {Bowler}, {Boyett}, {Cresci},
  {Curtis-Lake}, {D'Eugenio}, {Kumari}, {Looser}, {Maiolino}, {{\"U}bler},
  {Willott}, \& {Witstok}}]{kat23b}
{Katz}, H., {Saxena}, A., {Cameron}, A.~J., {et~al.} 2023{\natexlab{a}},
  \mnras, 518, 592

\bibitem[{{Katz} {et~al.}(2023{\natexlab{b}}){Katz}, {Saxena}, {Rosdahl},
  {Kimm}, {Blaizot}, {Garel}, {Michel-Dansac}, {Haehnelt}, {Ellis},
  {Penterrici}, {Devriendt}, \& {Slyz}}]{kat23a}
{Katz}, H., {Saxena}, A., {Rosdahl}, J., {et~al.} 2023{\natexlab{b}}, \mnras,
  518, 270

\bibitem[{{Kennicutt}(1998)}]{ken98}
{Kennicutt}, Robert~C., J. 1998, \apj, 498, 541

\bibitem[{{Kroupa}(2001)}]{kro01}
{Kroupa}, P. 2001, \mnras, 322, 231

\bibitem[{{Laporte} {et~al.}(2017){Laporte}, {Nakajima}, {Ellis}, {Zitrin},
  {Stark}, {Mainali}, \& {Roberts-Borsani}}]{lap17}
{Laporte}, N., {Nakajima}, K., {Ellis}, R.~S., {et~al.} 2017, \apj, 851, 40

\bibitem[{{Larson} {et~al.}(2022){Larson}, {Finkelstein}, {Hutchison},
  {Papovich}, {Bagley}, {Dickinson}, {Rojas-Ruiz}, {Ferguson}, {Jung},
  {Giavalisco}, {Grazian}, {Pentericci}, \& {Tacchella}}]{lar22}
{Larson}, R.~L., {Finkelstein}, S.~L., {Hutchison}, T.~A., {et~al.} 2022, \apj,
  930, 104

\bibitem[{{Leonova} {et~al.}(2022){Leonova}, {Oesch}, {Qin}, {Naidu}, {Wyithe},
  {de Barros}, {Bouwens}, {Ellis}, {Endsley}, {Hutter}, {Illingworth},
  {Kerutt}, {Labb{\'e}}, {Laporte}, {Magee}, {Mutch}, {Roberts-Borsani},
  {Smit}, {Stark}, {Stefanon}, {Tacchella}, \& {Zitrin}}]{leo22}
{Leonova}, E., {Oesch}, P.~A., {Qin}, Y., {et~al.} 2022, \mnras, 515, 5790

\bibitem[{{Luridiana} {et~al.}(2015){Luridiana}, {Morisset}, \& {Shaw}}]{pyneb}
{Luridiana}, V., {Morisset}, C., \& {Shaw}, R.~A. 2015, \aap, 573, A42

\bibitem[{{Madau} {et~al.}(1999){Madau}, {Haardt}, \& {Rees}}]{mad99}
{Madau}, P., {Haardt}, F., \& {Rees}, M.~J. 1999, \apj, 514, 648

\bibitem[{{Maiolino} {et~al.}(2015){Maiolino}, {Carniani}, {Fontana},
  {Vallini}, {Pentericci}, {Ferrara}, {Vanzella}, {Grazian}, {Gallerani},
  {Castellano}, {Cristiani}, {Brammer}, {Santini}, {Wagg}, \&
  {Williams}}]{mai15}
{Maiolino}, R., {Carniani}, S., {Fontana}, A., {et~al.} 2015, \mnras, 452, 54

\bibitem[{{Maji} {et~al.}(2022){Maji}, {Verhamme}, {Rosdahl}, {Garel},
  {Blaizot}, {Mauerhofer}, {Pittavino}, {Victoria Feser}, {Chuniaud}, {Kimm},
  {Katz}, \& {Haehnelt}}]{maj22}
{Maji}, M., {Verhamme}, A., {Rosdahl}, J., {et~al.} 2022, \aap, 663, A66

\bibitem[{{Malhotra} \& {Rhoads}(2006)}]{mal06}
{Malhotra}, S. \& {Rhoads}, J.~E. 2006, \apjl, 647, L95

\bibitem[{{Marchi} {et~al.}(2018){Marchi}, {Pentericci}, {Guaita}, {Schaerer},
  {Verhamme}, {Castellano}, {Ribeiro}, {Garilli}, {Le F{\`e}vre}, {Amorin},
  {Bardelli}, {Cassata}, {Durkalec}, {Grazian}, {Hathi}, {Lemaux}, {Maccagni},
  {Vanzella}, \& {Zucca}}]{mar18}
{Marchi}, F., {Pentericci}, L., {Guaita}, L., {et~al.} 2018, \aap, 614, A11

\bibitem[{{Mascia} {et~al.}(2023){Mascia}, {Pentericci}, {Calabr{\`o}}, {Treu},
  {Santini}, {Yang}, {Napolitano}, {Roberts-Borsani}, {Bergamini}, {Grillo},
  {Rosati}, {Vulcani}, {Castellano}, {Boyett}, {Fontana}, {Glazebrook},
  {Henry}, {Mason}, {Merlin}, {Morishita}, {Nanayakkara}, {Paris}, {Roy},
  {Williams}, {Wang}, {Brammer}, {Brada{\v{c}}}, {Chen}, {Kelly}, {Koekemoer},
  {Trenti}, \& {Windhorst}}]{mas23}
{Mascia}, S., {Pentericci}, L., {Calabr{\`o}}, A., {et~al.} 2023, \aap, 672,
  A155

\bibitem[{{Maseda} {et~al.}(2020){Maseda}, {Bacon}, {Lam}, {Matthee},
  {Brinchmann}, {Schaye}, {Labbe}, {Schmidt}, {Boogaard}, {Bouwens},
  {Cantalupo}, {Franx}, {Hashimoto}, {Inami}, {Kusakabe}, {Mahler},
  {Nanayakkara}, {Richard}, \& {Wisotzki}}]{mas20}
{Maseda}, M.~V., {Bacon}, R., {Lam}, D., {et~al.} 2020, \mnras, 493, 5120

\bibitem[{{Mason} {et~al.}(2019){Mason}, {Fontana}, {Treu}, {Schmidt}, {Hoag},
  {Abramson}, {Amorin}, {Brada{\v{c}}}, {Guaita}, {Jones}, {Henry}, {Malkan},
  {Pentericci}, {Trenti}, \& {Vanzella}}]{mas19a}
{Mason}, C.~A., {Fontana}, A., {Treu}, T., {et~al.} 2019, \mnras, 485, 3947

\bibitem[{{Mason} \& {Gronke}(2020)}]{mason20}
{Mason}, C.~A. \& {Gronke}, M. 2020, \mnras, 499, 1395

\bibitem[{{Mason} {et~al.}(2018){Mason}, {Treu}, {Dijkstra}, {Mesinger},
  {Trenti}, {Pentericci}, {de Barros}, \& {Vanzella}}]{mas18}
{Mason}, C.~A., {Treu}, T., {Dijkstra}, M., {et~al.} 2018, \apj, 856, 2

\bibitem[{{Matthee} {et~al.}(2022){Matthee}, {Naidu}, {Pezzulli}, {Gronke},
  {Sobral}, {Oesch}, {Hayes}, {Erb}, {Schaerer}, {Amor{\'\i}n}, {Tacchella},
  {Paulino-Afonso}, {Llerena}, {Calhau}, \& {R{\"o}ttgering}}]{mat22}
{Matthee}, J., {Naidu}, R.~P., {Pezzulli}, G., {et~al.} 2022, \mnras, 512, 5960

\bibitem[{{McLure} {et~al.}(2013){McLure}, {Dunlop}, {Bowler}, {Curtis-Lake},
  {Schenker}, {Ellis}, {Robertson}, {Koekemoer}, {Rogers}, {Ono}, {Ouchi},
  {Charlot}, {Wild}, {Stark}, {Furlanetto}, {Cirasuolo}, \& {Targett}}]{mcl13}
{McLure}, R.~J., {Dunlop}, J.~S., {Bowler}, R.~A.~A., {et~al.} 2013, \mnras,
  432, 2696

\bibitem[{{Mesinger} {et~al.}(2015){Mesinger}, {Aykutalp}, {Vanzella},
  {Pentericci}, {Ferrara}, \& {Dijkstra}}]{mes15}
{Mesinger}, A., {Aykutalp}, A., {Vanzella}, E., {et~al.} 2015, \mnras, 446, 566

\bibitem[{{Miralda-Escud{\'e}}(1998)}]{mir98}
{Miralda-Escud{\'e}}, J. 1998, \apj, 501, 15

\bibitem[{{Morales} {et~al.}(2021){Morales}, {Mason}, {Bruton}, {Gronke},
  {Haardt}, \& {Scarlata}}]{mor21}
{Morales}, A.~M., {Mason}, C.~A., {Bruton}, S., {et~al.} 2021, \apj, 919, 120

\bibitem[{{Naidu} {et~al.}(2022){Naidu}, {Matthee}, {Oesch}, {Conroy},
  {Sobral}, {Pezzulli}, {Hayes}, {Erb}, {Amor{\'\i}n}, {Gronke}, {Schaerer},
  {Tacchella}, {Kerutt}, {Paulino-Afonso}, {Calhau}, {Llerena}, \&
  {R{\"o}ttgering}}]{nai22}
{Naidu}, R.~P., {Matthee}, J., {Oesch}, P.~A., {et~al.} 2022, \mnras, 510, 4582

\bibitem[{{Nakajima} {et~al.}(2023){Nakajima}, {Ouchi}, {Isobe}, {Harikane},
  {Zhang}, {Ono}, {Umeda}, \& {Oguri}}]{nak23}
{Nakajima}, K., {Ouchi}, M., {Isobe}, Y., {et~al.} 2023, arXiv e-prints,
  arXiv:2301.12825

\bibitem[{{Ning} {et~al.}(2022){Ning}, {Jiang}, {Zheng}, \& {Wu}}]{nin22}
{Ning}, Y., {Jiang}, L., {Zheng}, Z.-Y., \& {Wu}, J. 2022, \apj, 926, 230

\bibitem[{{Oesch} {et~al.}(2015){Oesch}, {van Dokkum}, {Illingworth},
  {Bouwens}, {Momcheva}, {Holden}, {Roberts-Borsani}, {Smit}, {Franx},
  {Labb{\'e}}, {Gonz{\'a}lez}, \& {Magee}}]{oes15}
{Oesch}, P.~A., {van Dokkum}, P.~G., {Illingworth}, G.~D., {et~al.} 2015,
  \apjl, 804, L30

\bibitem[{{Oke} \& {Gunn}(1983)}]{oke83}
{Oke}, J.~B. \& {Gunn}, J.~E. 1983, \apj, 266, 713

\bibitem[{{Osterbrock}(1989)}]{ost89}
{Osterbrock}, D.~E. 1989, {Astrophysics of gaseous nebulae and active galactic
  nuclei}

\bibitem[{{Osterbrock} \& {Ferland}(2006)}]{ost06}
{Osterbrock}, D.~E. \& {Ferland}, G.~J. 2006, {Astrophysics of gaseous nebulae
  and active galactic nuclei}

\bibitem[{{Paalvast} {et~al.}(2018){Paalvast}, {Verhamme}, {Straka},
  {Brinchmann}, {Herenz}, {Carton}, {Gunawardhana}, {Boogaard}, {Cantalupo},
  {Contini}, {Epinat}, {Inami}, {Marino}, {Maseda}, {Michel-Dansac}, {Muzahid},
  {Nanayakkara}, {Pezzulli}, {Richard}, {Schaye}, {Segers}, {Urrutia}, {Wendt},
  \& {Wisotzki}}]{paa18}
{Paalvast}, M., {Verhamme}, A., {Straka}, L.~A., {et~al.} 2018, \aap, 618, A40

\bibitem[{{Pentericci} {et~al.}(2016){Pentericci}, {Carniani}, {Castellano},
  {Fontana}, {Maiolino}, {Guaita}, {Vanzella}, {Grazian}, {Santini}, {Yan},
  {Cristiani}, {Conselice}, {Giavalisco}, {Hathi}, \& {Koekemoer}}]{pen16}
{Pentericci}, L., {Carniani}, S., {Castellano}, M., {et~al.} 2016, \apjl, 829,
  L11

\bibitem[{{Pentericci} {et~al.}(2011){Pentericci}, {Fontana}, {Vanzella},
  {Castellano}, {Grazian}, {Dijkstra}, {Boutsia}, {Cristiani}, {Dickinson},
  {Giallongo}, {Giavalisco}, {Maiolino}, {Moorwood}, {Paris}, \&
  {Santini}}]{pen11}
{Pentericci}, L., {Fontana}, A., {Vanzella}, E., {et~al.} 2011, \apj, 743, 132

\bibitem[{{Pentericci} {et~al.}(2014){Pentericci}, {Vanzella}, {Fontana},
  {Castellano}, {Treu}, {Mesinger}, {Dijkstra}, {Grazian}, {Brada{\v{c}}},
  {Conselice}, {Cristiani}, {Dunlop}, {Galametz}, {Giavalisco}, {Giallongo},
  {Koekemoer}, {McLure}, {Maiolino}, {Paris}, \& {Santini}}]{pen14}
{Pentericci}, L., {Vanzella}, E., {Fontana}, A., {et~al.} 2014, \apj, 793, 113

\bibitem[{{Planck Collaboration} {et~al.}(2020){Planck Collaboration},
  {Aghanim}, {Akrami}, {Ashdown}, {Aumont}, {Baccigalupi}, {Ballardini},
  {Banday}, {Barreiro}, {Bartolo}, {Basak}, {Battye}, {Benabed}, {Bernard},
  {Bersanelli}, {Bielewicz}, {Bock}, {Bond}, {Borrill}, {Bouchet}, {Boulanger},
  {Bucher}, {Burigana}, {Butler}, {Calabrese}, {Cardoso}, {Carron},
  {Challinor}, {Chiang}, {Chluba}, {Colombo}, {Combet}, {Contreras}, {Crill},
  {Cuttaia}, {de Bernardis}, {de Zotti}, {Delabrouille}, {Delouis}, {Di
  Valentino}, {Diego}, {Dor{\'e}}, {Douspis}, {Ducout}, {Dupac}, {Dusini},
  {Efstathiou}, {Elsner}, {En{\ss}lin}, {Eriksen}, {Fantaye}, {Farhang},
  {Fergusson}, {Fernandez-Cobos}, {Finelli}, {Forastieri}, {Frailis},
  {Fraisse}, {Franceschi}, {Frolov}, {Galeotta}, {Galli}, {Ganga},
  {G{\'e}nova-Santos}, {Gerbino}, {Ghosh}, {Gonz{\'a}lez-Nuevo}, {G{\'o}rski},
  {Gratton}, {Gruppuso}, {Gudmundsson}, {Hamann}, {Handley}, {Hansen},
  {Herranz}, {Hildebrandt}, {Hivon}, {Huang}, {Jaffe}, {Jones}, {Karakci},
  {Keih{\"a}nen}, {Keskitalo}, {Kiiveri}, {Kim}, {Kisner}, {Knox},
  {Krachmalnicoff}, {Kunz}, {Kurki-Suonio}, {Lagache}, {Lamarre}, {Lasenby},
  {Lattanzi}, {Lawrence}, {Le Jeune}, {Lemos}, {Lesgourgues}, {Levrier},
  {Lewis}, {Liguori}, {Lilje}, {Lilley}, {Lindholm}, {L{\'o}pez-Caniego},
  {Lubin}, {Ma}, {Mac{\'\i}as-P{\'e}rez}, {Maggio}, {Maino}, {Mandolesi},
  {Mangilli}, {Marcos-Caballero}, {Maris}, {Martin}, {Martinelli},
  {Mart{\'\i}nez-Gonz{\'a}lez}, {Matarrese}, {Mauri}, {McEwen}, {Meinhold},
  {Melchiorri}, {Mennella}, {Migliaccio}, {Millea}, {Mitra},
  {Miville-Desch{\^e}nes}, {Molinari}, {Montier}, {Morgante}, {Moss}, {Natoli},
  {N{\o}rgaard-Nielsen}, {Pagano}, {Paoletti}, {Partridge}, {Patanchon},
  {Peiris}, {Perrotta}, {Pettorino}, {Piacentini}, {Polastri}, {Polenta},
  {Puget}, {Rachen}, {Reinecke}, {Remazeilles}, {Renzi}, {Rocha}, {Rosset},
  {Roudier}, {Rubi{\~n}o-Mart{\'\i}n}, {Ruiz-Granados}, {Salvati}, {Sandri},
  {Savelainen}, {Scott}, {Shellard}, {Sirignano}, {Sirri}, {Spencer},
  {Sunyaev}, {Suur-Uski}, {Tauber}, {Tavagnacco}, {Tenti}, {Toffolatti},
  {Tomasi}, {Trombetti}, {Valenziano}, {Valiviita}, {Van Tent}, {Vibert},
  {Vielva}, {Villa}, {Vittorio}, {Wandelt}, {Wehus}, {White}, {White},
  {Zacchei}, \& {Zonca}}]{planck}
{Planck Collaboration}, {Aghanim}, N., {Akrami}, Y., {et~al.} 2020, \aap, 641,
  A6

\bibitem[{{Qin} {et~al.}(2022){Qin}, {Wyithe}, {Oesch}, {Illingworth},
  {Leonova}, {Mutch}, \& {Naidu}}]{qin22}
{Qin}, Y., {Wyithe}, J. S.~B., {Oesch}, P.~A., {et~al.} 2022, \mnras, 510, 3858

\bibitem[{{Reddy} {et~al.}(2022){Reddy}, {Topping}, {Shapley}, {Steidel},
  {Sanders}, {Du}, {Coil}, {Mobasher}, {Price}, \& {Shivaei}}]{red22}
{Reddy}, N.~A., {Topping}, M.~W., {Shapley}, A.~E., {et~al.} 2022, \apj, 926,
  31

\bibitem[{{Rieke} {et~al.}(2023){Rieke}, {Kelly}, {Misselt}, {Stansberry},
  {Boyer}, {Beatty}, {Egami}, {Florian}, {Greene}, {Hainline}, {Leisenring},
  {Roellig}, {Schlawin}, {Sun}, {Tinnin}, {Williams}, {Willmer}, {Wilson},
  {Clark}, {Rohrbach}, {Brooks}, {Canipe}, {Correnti}, {DiFelice}, {Gennaro},
  {Girard}, {Hartig}, {Hilbert}, {Koekemoer}, {Nikolov}, {Pirzkal}, {Rest},
  {Robberto}, {Sunnquist}, {Telfer}, {Wu}, {Ferry}, {Lewis}, {Baum},
  {Beichman}, {Doyon}, {Dressler}, {Eisenstein}, {Ferrarese}, {Hodapp},
  {Horner}, {Jaffe}, {Johnstone}, {Krist}, {Martin}, {McCarthy}, {Meyer},
  {Rieke}, {Trauger}, \& {Young}}]{rie23a}
{Rieke}, M.~J., {Kelly}, D.~M., {Misselt}, K., {et~al.} 2023, \pasp, 135,
  028001

\bibitem[{{Roberts-Borsani} {et~al.}(2023){Roberts-Borsani}, {Treu}, {Mason},
  {Ellis}, {Laporte}, {Schmidt}, {Bradac}, {Fontana}, {Morishita}, \&
  {Santini}}]{roberts23}
{Roberts-Borsani}, G., {Treu}, T., {Mason}, C., {et~al.} 2023, \apj, 948, 54

\bibitem[{{Robertson}(2022)}]{rob22a}
{Robertson}, B.~E. 2022, \araa, 60, 121

\bibitem[{{Robertson} {et~al.}(2015){Robertson}, {Ellis}, {Furlanetto}, \&
  {Dunlop}}]{rob15}
{Robertson}, B.~E., {Ellis}, R.~S., {Furlanetto}, S.~R., \& {Dunlop}, J.~S.
  2015, \apjl, 802, L19

\bibitem[{{Robertson} {et~al.}(2013){Robertson}, {Furlanetto}, {Schneider},
  {Charlot}, {Ellis}, {Stark}, {McLure}, {Dunlop}, {Koekemoer}, {Schenker},
  {Ouchi}, {Ono}, {Curtis-Lake}, {Rogers}, {Bowler}, \& {Cirasuolo}}]{rob13}
{Robertson}, B.~E., {Furlanetto}, S.~R., {Schneider}, E., {et~al.} 2013, \apj,
  768, 71

\bibitem[{{Robertson} {et~al.}(2023){Robertson}, {Tacchella}, {Johnson},
  {Hainline}, {Whitler}, {Eisenstein}, {Endsley}, {Rieke}, {Stark}, {Alberts},
  {Dressler}, {Egami}, {Hausen}, {Rieke}, {Shivaei}, {Williams}, {Willmer},
  {Arribas}, {Bonaventura}, {Bunker}, {Cameron}, {Carniani}, {Charlot},
  {Chevallard}, {Curti}, {Curtis-Lake}, {D'Eugenio}, {Jakobsen}, {Looser},
  {L{\"u}tzgendorf}, {Maiolino}, {Maseda}, {Rawle}, {Rix}, {Smit}, {{\"U}bler},
  {Willott}, {Witstok}, {Baum}, {Bhatawdekar}, {Boyett}, {Chen}, {de Graaff},
  {Florian}, {Helton}, {Hviding}, {Ji}, {Kumari}, {Lyu}, {Nelson}, {Sandles},
  {Saxena}, {Suess}, {Sun}, {Topping}, \& {Wallace}}]{rob23}
{Robertson}, B.~E., {Tacchella}, S., {Johnson}, B.~D., {et~al.} 2023, Nature
  Astronomy, 7, 611

\bibitem[{{Saldana-Lopez} {et~al.}(2023){Saldana-Lopez}, {Schaerer},
  {Chisholm}, {Calabr{\`o}}, {Pentericci}, {Cullen}, {Saxena}, {Amor{\'\i}n},
  {Carnall}, {Fontanot}, {Fynbo}, {Guaita}, {Hathi}, {Hibon}, {Ji}, {McLeod},
  {Pompei}, \& {Zamorani}}]{sal23}
{Saldana-Lopez}, A., {Schaerer}, D., {Chisholm}, J., {et~al.} 2023, \mnras,
  522, 6295

\bibitem[{{Salim} {et~al.}(2018){Salim}, {Boquien}, \& {Lee}}]{sal18}
{Salim}, S., {Boquien}, M., \& {Lee}, J.~C. 2018, \apj, 859, 11

\bibitem[{{Salpeter}(1955)}]{sal55}
{Salpeter}, E.~E. 1955, \apj, 121, 161

\bibitem[{{Sanders} {et~al.}(2023){Sanders}, {Shapley}, {Topping}, {Reddy}, \&
  {Brammer}}]{san23}
{Sanders}, R.~L., {Shapley}, A.~E., {Topping}, M.~W., {Reddy}, N.~A., \&
  {Brammer}, G.~B. 2023, arXiv e-prints, arXiv:2301.06696

\bibitem[{{Saxena} {et~al.}(2023){Saxena}, {Bunker}, {Jones}, {Stark},
  {Cameron}, {Witstok}, {Arribas}, {Baker}, {Baum}, {Bhatawdekar}, {Bowler},
  {Boyett}, {Carniani}, {Charlot}, {Chevallard}, {Curti}, {Curtis-Lake},
  {Eisenstein}, {Endsley}, {Hainline}, {Helton}, {Johnson}, {Kumari}, {Looser},
  {Maiolino}, {Rieke}, {Rix}, {Robertson}, {Sandles}, {Simmonds}, {Smit},
  {Tacchella}, {Williams}, {Willmer}, \& {Willott}}]{sax23b}
{Saxena}, A., {Bunker}, A.~J., {Jones}, G.~C., {et~al.} 2023, arXiv e-prints,
  arXiv:2306.04536

\bibitem[{{Saxena} {et~al.}(2022{\natexlab{a}}){Saxena}, {Cryer}, {Ellis},
  {Pentericci}, {Calabr{\`o}}, {Mascia}, {Saldana-Lopez}, {Schaerer}, {Katz},
  {Llerena}, \& {Amor{\'\i}n}}]{sax22b}
{Saxena}, A., {Cryer}, E., {Ellis}, R.~S., {et~al.} 2022{\natexlab{a}}, \mnras,
  517, 1098

\bibitem[{{Saxena} {et~al.}(2022{\natexlab{b}}){Saxena}, {Pentericci}, {Ellis},
  {Guaita}, {Calabr{\`o}}, {Schaerer}, {Vanzella}, {Amor{\'\i}n}, {Bolzonella},
  {Castellano}, {Fontanot}, {Hathi}, {Hibon}, {Llerena}, {Mannucci},
  {Saldana-Lopez}, {Talia}, \& {Zamorani}}]{sax22a}
{Saxena}, A., {Pentericci}, L., {Ellis}, R.~S., {et~al.} 2022{\natexlab{b}},
  \mnras, 511, 120

\bibitem[{{Schaerer} {et~al.}(2022){Schaerer}, {Izotov}, {Worseck}, {Berg},
  {Chisholm}, {Jaskot}, {Nakajima}, {Ravindranath}, {Thuan}, \&
  {Verhamme}}]{sch22a}
{Schaerer}, D., {Izotov}, Y.~I., {Worseck}, G., {et~al.} 2022, \aap, 658, L11

\bibitem[{{Schenker} {et~al.}(2013){Schenker}, {Robertson}, {Ellis}, {Ono},
  {McLure}, {Dunlop}, {Koekemoer}, {Bowler}, {Ouchi}, {Curtis-Lake}, {Rogers},
  {Schneider}, {Charlot}, {Stark}, {Furlanetto}, \& {Cirasuolo}}]{sch13}
{Schenker}, M.~A., {Robertson}, B.~E., {Ellis}, R.~S., {et~al.} 2013, \apj,
  768, 196

\bibitem[{{Shapley} {et~al.}(2023){Shapley}, {Sanders}, {Reddy}, {Topping}, \&
  {Brammer}}]{sha23}
{Shapley}, A.~E., {Sanders}, R.~L., {Reddy}, N.~A., {Topping}, M.~W., \&
  {Brammer}, G.~B. 2023, arXiv e-prints, arXiv:2301.03241

\bibitem[{{Shivaei} {et~al.}(2020){Shivaei}, {Reddy}, {Rieke}, {Shapley},
  {Kriek}, {Battisti}, {Mobasher}, {Sanders}, {Fetherolf}, {Azadi}, {Coil},
  {Freeman}, {de Groot}, {Leung}, {Price}, {Siana}, \& {Zick}}]{shi20}
{Shivaei}, I., {Reddy}, N., {Rieke}, G., {et~al.} 2020, \apj, 899, 117

\bibitem[{{Simmonds} {et~al.}(2023){Simmonds}, {Tacchella}, {Maseda},
  {Williams}, {Baker}, {Witten}, {Johnson}, {Robertson}, {Saxena}, {Sun},
  {Witstok}, {Bhatawdekar}, {Boyett}, {Bunker}, {Charlot}, {Curtis-Lake},
  {Egami}, {Eisenstein}, {Ji}, {Maiolino}, {Sandles}, {Smit}, {{\"U}bler}, \&
  {Willott}}]{sim23}
{Simmonds}, C., {Tacchella}, S., {Maseda}, M., {et~al.} 2023, \mnras, 523, 5468

\bibitem[{{Stark} {et~al.}(2017){Stark}, {Ellis}, {Charlot}, {Chevallard},
  {Tang}, {Belli}, {Zitrin}, {Mainali}, {Gutkin}, {Vidal-Garc{\'\i}a},
  {Bouwens}, \& {Oesch}}]{sta17}
{Stark}, D.~P., {Ellis}, R.~S., {Charlot}, S., {et~al.} 2017, \mnras, 464, 469

\bibitem[{{Steidel} {et~al.}(1996){Steidel}, {Giavalisco}, {Dickinson}, \&
  {Adelberger}}]{ste96}
{Steidel}, C.~C., {Giavalisco}, M., {Dickinson}, M., \& {Adelberger}, K.~L.
  1996, \aj, 112, 352

\bibitem[{{Sun} {et~al.}(2023){Sun}, {Egami}, {Pirzkal}, {Rieke}, {Baum},
  {Boyer}, {Boyett}, {Bunker}, {Cameron}, {Curti}, {Eisenstein}, {Gennaro},
  {Greene}, {Jaffe}, {Kelly}, {Koekemoer}, {Kumari}, {Maiolino}, {Maseda},
  {Perna}, {Rest}, {Robertson}, {Schlawin}, {Smit}, {Stansberry}, {Sunnquist},
  {Tacchella}, {Williams}, \& {Willmer}}]{sun23}
{Sun}, F., {Egami}, E., {Pirzkal}, N., {et~al.} 2023, \apj, 953, 53

\bibitem[{{Tacchella} {et~al.}(2023{\natexlab{a}}){Tacchella}, {Eisenstein},
  {Hainline}, {Johnson}, {Baker}, {Helton}, {Robertson}, {Suess}, {Chen},
  {Nelson}, {Pusk{\'a}s}, {Sun}, {Alberts}, {Egami}, {Hausen}, {Rieke},
  {Rieke}, {Shivaei}, {Williams}, {Willmer}, {Bunker}, {Cameron}, {Carniani},
  {Charlot}, {Curti}, {Curtis-Lake}, {Looser}, {Maiolino}, {Maseda}, {Rawle},
  {Rix}, {Smit}, {{\"U}bler}, {Willott}, {Witstok}, {Baum}, {Bhatawdekar},
  {Boyett}, {Danhaive}, {de Graaff}, {Endsley}, {Ji}, {Lyu}, {Sandles},
  {Saxena}, {Scholtz}, {Topping}, \& {Whitler}}]{tac23b}
{Tacchella}, S., {Eisenstein}, D.~J., {Hainline}, K., {et~al.}
  2023{\natexlab{a}}, \apj, 952, 74

\bibitem[{{Tacchella} {et~al.}(2023{\natexlab{b}}){Tacchella}, {Johnson},
  {Robertson}, {Carniani}, {D'Eugenio}, {Kumari}, {Maiolino}, {Nelson},
  {Suess}, {{\"U}bler}, {Williams}, {Adebusola}, {Alberts}, {Arribas},
  {Bhatawdekar}, {Bonaventura}, {Bowler}, {Bunker}, {Cameron}, {Curti},
  {Egami}, {Eisenstein}, {Frye}, {Hainline}, {Helton}, {Ji}, {Looser}, {Lyu},
  {Perna}, {Rawle}, {Rieke}, {Rieke}, {Saxena}, {Sandles}, {Shivaei},
  {Simmonds}, {Sun}, {Willmer}, {Willott}, \& {Witstok}}]{tac23a}
{Tacchella}, S., {Johnson}, B.~D., {Robertson}, B.~E., {et~al.}
  2023{\natexlab{b}}, \mnras, 522, 6236

\bibitem[{{Tang} {et~al.}(2023){Tang}, {Stark}, {Chen}, {Mason}, {Topping},
  {Endsley}, {Senchyna}, {Plat}, {Lu}, {Whitler}, {Robertson}, \&
  {Charlot}}]{tan23}
{Tang}, M., {Stark}, D.~P., {Chen}, Z., {et~al.} 2023, arXiv e-prints,
  arXiv:2301.07072

\bibitem[{{Tilvi} {et~al.}(2020){Tilvi}, {Malhotra}, {Rhoads}, {Coughlin},
  {Zheng}, {Finkelstein}, {Veilleux}, {Mobasher}, {Wang}, {Probst}, {Swaters},
  {Hibon}, {Joshi}, {Zabl}, {Jiang}, {Pharo}, \& {Yang}}]{til20}
{Tilvi}, V., {Malhotra}, S., {Rhoads}, J.~E., {et~al.} 2020, \apjl, 891, L10

\bibitem[{{Topping} {et~al.}(2022){Topping}, {Stark}, {Endsley}, {Plat},
  {Whitler}, {Chen}, \& {Charlot}}]{top22}
{Topping}, M.~W., {Stark}, D.~P., {Endsley}, R., {et~al.} 2022, \apj, 941, 153

\bibitem[{{Trapp} {et~al.}(2023){Trapp}, {Furlanetto}, \& {Davies}}]{tra23}
{Trapp}, A.~C., {Furlanetto}, S.~R., \& {Davies}, F.~B. 2023, \mnras, 524, 5891

\bibitem[{{Trump} {et~al.}(2023){Trump}, {Arrabal Haro}, {Simons}, {Backhaus},
  {Amor{\'\i}n}, {Dickinson}, {Fern{\'a}ndez}, {Papovich}, {Nicholls},
  {Kewley}, {Brunker}, {Salzer}, {Wilkins}, {Almaini}, {Bagley}, {Berg},
  {Bhatawdekar}, {Bisigello}, {Buat}, {Burgarella}, {Calabr{\`o}}, {Casey},
  {Ciesla}, {Cleri}, {Cole}, {Cooper}, {Cooray}, {Costantin}, {Croton},
  {Ferguson}, {Finkelstein}, {Fujimoto}, {Gardner}, {Gawiser}, {Giavalisco},
  {Grazian}, {Grogin}, {Hathi}, {Hirschmann}, {Holwerda}, {Huertas-Company},
  {Hutchison}, {Jogee}, {Juneau}, {Jung}, {Kartaltepe}, {Kirkpatrick},
  {Kocevski}, {Koekemoer}, {Lotz}, {Lucas}, {Magnelli}, {Matharu},
  {P{\'e}rez-Gonz{\'a}lez}, {Pirzkal}, {Rafelski}, {Rose}, {Seill{\'e}},
  {Somerville}, {Straughn}, {Tacchella}, {Vanderhoof}, {Weiner}, {Wuyts},
  {Yung}, \& {Zavala}}]{tru23}
{Trump}, J.~R., {Arrabal Haro}, P., {Simons}, R.~C., {et~al.} 2023, \apj, 945,
  35

\bibitem[{{Valentino} {et~al.}(2022){Valentino}, {Brammer}, {Fujimoto},
  {Heintz}, {Weaver}, {Strait}, {Gould}, {Mason}, {Watson}, {Laursen}, \&
  {Toft}}]{val22}
{Valentino}, F., {Brammer}, G., {Fujimoto}, S., {et~al.} 2022, \apjl, 929, L9

\bibitem[{{Vanzella} {et~al.}(2011){Vanzella}, {Pentericci}, {Fontana},
  {Grazian}, {Castellano}, {Boutsia}, {Cristiani}, {Dickinson}, {Gallozzi},
  {Giallongo}, {Giavalisco}, {Maiolino}, {Moorwood}, {Paris}, \&
  {Santini}}]{van11}
{Vanzella}, E., {Pentericci}, L., {Fontana}, A., {et~al.} 2011, \apjl, 730, L35

\bibitem[{{Verhamme} {et~al.}(2015){Verhamme}, {Orlitov{\'a}}, {Schaerer}, \&
  {Hayes}}]{ver15}
{Verhamme}, A., {Orlitov{\'a}}, I., {Schaerer}, D., \& {Hayes}, M. 2015, \aap,
  578, A7

\bibitem[{{Verhamme} {et~al.}(2017){Verhamme}, {Orlitov{\'a}}, {Schaerer},
  {Izotov}, {Worseck}, {Thuan}, \& {Guseva}}]{ver17}
{Verhamme}, A., {Orlitov{\'a}}, I., {Schaerer}, D., {et~al.} 2017, \aap, 597,
  A13

\bibitem[{{Whitler} {et~al.}(2020){Whitler}, {Mason}, {Ren}, {Dijkstra},
  {Mesinger}, {Pentericci}, {Trenti}, \& {Treu}}]{whi20}
{Whitler}, L.~R., {Mason}, C.~A., {Ren}, K., {et~al.} 2020, \mnras, 495, 3602

\bibitem[{{Witstok} {et~al.}(2023){Witstok}, {Smit}, {Saxena}, {Jones},
  {Helton}, {Sun}, {Maiolino}, {Stark}, {Bunker}, {Arribas}, {Baker},
  {Bhatawdekar}, {Boyett}, {Cameron}, {Carniani}, {Charlot}, {Chevallard},
  {Curti}, {Curtis-Lake}, {Eisenstein}, {Endsley}, {Hainline}, {Ji}, {Johnson},
  {Kumari}, {Looser}, {Nelson}, {Perna}, {Rix}, {Robertson}, {Sandles},
  {Scholtz}, {Simmonds}, {Tacchella}, {{\"U}bler}, {Williams}, {Willmer}, \&
  {Willott}}]{wit23b}
{Witstok}, J., {Smit}, R., {Saxena}, A., {et~al.} 2023, arXiv e-prints,
  arXiv:2306.04627

\bibitem[{{Zackrisson} {et~al.}(2013){Zackrisson}, {Inoue}, \&
  {Jensen}}]{zac13}
{Zackrisson}, E., {Inoue}, A.~K., \& {Jensen}, H. 2013, \apj, 777, 39

\bibitem[{{Zitrin} {et~al.}(2015){Zitrin}, {Labb{\'e}}, {Belli}, {Bouwens},
  {Ellis}, {Roberts-Borsani}, {Stark}, {Oesch}, \& {Smit}}]{zit15}
{Zitrin}, A., {Labb{\'e}}, I., {Belli}, S., {et~al.} 2015, \apjl, 810, L12

\end{thebibliography}

\end{document}